\documentclass[final]{article}

\input glyphtounicode
\usepackage[T1]{fontenc}
\usepackage[utf8]{inputenc}

\usepackage{ifdraft}

\usepackage[english]{babel}

\usepackage{mathtools}
\usepackage{enumitem}
\usepackage[amsmath,thmmarks,hyperref]{ntheorem}

\usepackage{xcolor}
\definecolor{hypercolor}{rgb}{0,0.2,0.7}

\usepackage{footmisc}

\usepackage[numbers,sort&compress]{natbib}
\usepackage[final]{graphicx}
\usepackage{tikz}
\usetikzlibrary{calc,decorations.markings}

\newif\iffancyfont%
\fancyfontfalse%

\IfFileExists{MinionPro.sty}{\IfFileExists{MyriadPro.sty}{\fancyfonttrue}{}}{}

\iffancyfont%
  \usepackage[lf,medfamily,italicgreek,openg]{MinionPro}
  \usepackage[lf,medfamily,onlytext,scale=0.96]{MyriadPro}

  \renewcommand{\dotsm}{{\mathinner{\cdotp\cdotp\cdotp}}}

  \DeclareRobustCommand{\defn}{\mathrel{{\vdotdot}{\equal}}}

  \DeclareMathSymbol{\widehatsym}{\mathord}{largesymbols}{'302}

  \usepackage[toc,eqno,bib]{tabfigures}
  \newtagform{tabular}[\figureversion{tabular}]{(}{)}
  \usetagform{tabular}
\else%
  \usepackage[charter,greekuppercase=italicized,cal=cmcal]{mathdesign}

  \let\uppi\piup
  \let\upDelta\Deltaup

  \newcommand{\defn}{\coloneqq}

  \DeclareMathSymbol{\widehatsym}{\mathord}{largesymbols}{"62}

  \let\righthalfcup\lrcorner
\fi%

\usepackage{dsfont}

\usepackage[scr=rsfso]{mathalfa}

\ifdraft{
  \let\what\widehat

}{
  \usepackage{scalerel}
  \usepackage{accents}
  \newcommand\lowerwidehatsym{%
    \text{\smash{\kern-.1ex\raisebox{-1.3ex}{%
      $\widehatsym$}}}}
  \newcommand\what[1]{%
    \mathchoice%
      {\accentset{\displaystyle\ignoremathstyle\hstretch{0.5}\lowerwidehatsym}{#1}}%
      {\accentset{\textstyle\ignoremathstyle\hstretch{0.5}\lowerwidehatsym}{#1}}%
      {\accentset{\scriptstyle\ignoremathstyle\hstretch{0.5}\lowerwidehatsym}{#1}}%
      {\accentset{\scriptscriptstyle\ignoremathstyle\hstretch{0.5}\lowerwidehatsym}{#1}}%
  }
  
}

\usepackage{microtype}

\usepackage[sf,bf,small]{titlesec}
\usepackage[labelfont={sf,bf},labelsep=period]{caption}

\setlist[1]{labelindent=\parindent}
\setlist[description]{font=\sffamily\bfseries,align=right,labelsep=1em}

\numberwithin{equation}{section}

\makeatletter
\newcommand{\subtitle}[1]{\newcommand{\@subtitle}{#1}}
\newcounter{and}
\newdimen{\instindent}
\newcommand{\institute}[1]{\newcommand{\@institute}{#1}}
\newcommand{\inst}[1]{\unskip\smash{$^{#1}$}\setcounter{and}{1}\ignorespaces}
\newcommand{\email}[1]{\href{mailto:#1}{#1}}
\renewcommand{\maketitle}{
  {% title
    \raggedright%
    \LARGE%
    \noindent%
    \bfseries%
    \sffamily%
    \@title%
    \newline
    \Large
    \@subtitle
    \par
  }

  \vspace{1.5\baselineskip}

  {% author
    \raggedright%
    \renewcommand{\and}{\unskip, \ignorespaces}%
    \noindent\ignorespaces\@author\par
  }

  \vspace{0.5\baselineskip}

  {% institute
    \small%
    \parindent=0pt%
    \parskip=0pt%
    \setcounter{and}{1}%
    \renewcommand{\and}{%
      \par\stepcounter{and}%
      \hangindent\instindent%
      \noindent%
      \hbox to \instindent{\hss\smash{$^{\theand}$\enspace}}\ignorespaces%
    }%
    \setbox0=\vbox{\@institute}%
    \ifnum\value{and}>9\relax\setbox0=\hbox{$^{88}$\enspace}%
    \else\setbox0=\hbox{$^{8}$\enspace}\fi%
    \instindent=\wd0\relax%
    \ifnum\value{and}=1\relax%
    \else%
      \setcounter{and}{1}%
      \hangindent\instindent%
      \noindent%
      \hbox to \instindent{\hss\smash{$^{\theand}$}\enspace}\ignorespaces%
    \fi%
    \ignorespaces%
    \@institute\par
  }
}
\makeatother

\renewenvironment{abstract}{
  \addvspace{1.5\baselineskip}%
  \topsep=0pt\partopsep=0pt%
  \trivlist\item[\hspace{\labelsep}\bfseries\sffamily Abstract.]
}{}

\newenvironment{acknowledgments}{
  \addvspace{1.5\baselineskip}%
  \topsep=0pt\partopsep=0pt%
  \trivlist\item[\hspace{\labelsep}\bfseries\sffamily Acknowledgments.]
}{}

\usepackage[obeyDraft]{todonotes}

\ifdraft{
  \usepackage{soul}\setstcolor{red}
  \colorlet{lightblue}{cyan!60}

  \usepackage[notref,notcite]{showkeys}

  \usepackage{filemod}
  \usepackage{fancyhdr}
  \pagestyle{fancy}
  \fancyhf{}
  
  \fancyhead[C]{Draft version. Last modified: \filemodprint{\jobname}}
  \fancyfoot[C]{\thepage}
}{
  \newcommand{\st}[1]{\csname @latex@warning\endcsname{Use of \protect\st \space in final-mode detected}\ignorespaces}
  \newcommand{\hl}[1]{\csname @latex@warning\endcsname{Use of \protect\hl \space in final-mode detected}#1}
  
}

\newcommand\latinabbr\textit
\newcommand{\ie}{\latinabbr{i.e.}}
\newcommand{\eg}{\latinabbr{e.g.}}
\newcommand{\viz}{\latinabbr{viz.}}

\newcommand{\ia}{\latinabbr{i.a.}}

\newcommand{\e}{\mathrm{e}}
\newcommand{\im}{\mathrm{i}}

\newcommand{\field}[1][K]{{\mathbb{#1}}}

\newcommand{\RR}{{\field[R]}}
\newcommand{\CC}{{\field[C]}}

\DeclareMathOperator{\adj}{adj}

\renewcommand{\Re}{\operatorname{Re}}

\newcommand{\WF}{\mathrm{WF}}

\newcommand{\dif}{\mathrm{d}}

\renewcommand{\div}{\operatorname{div}}

\newcommand{\one}{\mathds{1}}

\let\conj\overline%
\DeclarePairedDelimiter{\abs}{\lvert}{\rvert}
\DeclarePairedDelimiter{\norm}{\lVert}{\rVert}
\newcommand{\norder}[1]{\mskip2mu\mathord{:}\mskip1mu{#1}\mskip1mu\mathord{:}\mskip2mu} % Normal order

\newcommand{\conlaw}{\mathop{\#}\nolimits}

\newcommand{\iprod}[1]{#1 \mathbin{\righthalfcup}}

\DeclarePairedDelimiterX\ip[2]{\langle}{\rangle}{#1 \,\delimsize\vert\, #2}

\newcommand{\hyp}{-\penalty0\hskip0pt\relax}

\newcommand{\N}{\aleph}

\theoremnumbering{arabic}
\qedsymbol{\ensuremath{\quad\Box}}

\theoremstyle{plain}
\theorembodyfont{\itshape}
\theoremheaderfont{\normalfont\sffamily\bfseries}
\theoremseparator{.}

\theorembodyfont{\upshape}

\theoremstyle{nonumberplain}
\theorembodyfont{\normalfont}
\theoremheaderfont{\normalfont\sffamily\bfseries}
\theoremsymbol{\ensuremath{\quad\Box}}

\title{Quantum Energy Inequalities in Pre-Metric Electrodynamics}
\subtitle{}

\author{
  Christopher J.\ Fewster\inst{1}
  \and
  Christian Pfeifer\inst{2}
  \and
  Daniel Siemssen\inst{3,4}
}

\institute{
  Department of Mathematics, University of York, Heslington, York YO10 5DD, United Kingdom.\\
  E-mail:~\email{chris.fewster@york.ac.uk}
  \and
  Laboratory of Theoretical Physics, Institute of Physics, University of Tartu, W. Ostwaldi 1, 50411 Tartu, Estonia.
  E-mail:~\email{christian.pfeifer@ut.ee}
  \and
  Department of Mathematical Methods in Physics, Faculty of Physics, University of Warsaw, Pasteura 5, 02-093 Warszawa, Poland.
  \and
  Department of Mathematics and Informatics, University of Wuppertal, Gaußstraße 20, 42119 Wuppertal, Germany.
  E-mail:~\email{siemssen@uni-wuppertal.de}
}

\usepackage[breaklinks,final]{hyperref}
\hypersetup{
  unicode,
  colorlinks,
  linkcolor=hypercolor,
  urlcolor=hypercolor,
  citecolor=hypercolor,
  bookmarksnumbered,
  bookmarksdepth=2,
  pdfauthor={Chris Fewster, Christian Pfeifer, Daniel Siemssen},
  pdftitle={Quantum Energy Inequalities in Pre-Metric Electrodynamics}
}

\begin{document}

\maketitle

\begin{abstract}
  Pre-metric electrodynamics is a covariant framework for electromagnetism with a general constitutive law.
  Its lightcone structure can be more complicated than that of Maxwell theory as is shown by the phenomenon of birefringence.
  We study the energy density of quantized pre-metric electrodynamics theories with linear constitutive laws admitting a single hyperbolicity double-cone and show that averages of the energy density along the worldlines of suitable observers obey a Quantum Energy Inequality (QEI) in states that satisfy a microlocal spectrum condition.
  The worldlines must meet two conditions: (a) the classical weak energy condition must hold along them, and (b) their velocity vectors have positive contractions with all positive frequency null covectors (we call such trajectories `subluminal').

  After stating our general results, we explicitly quantize the electromagnetic potential in a translationally invariant uniaxial birefringent crystal.
  Since the propagation of light in such a crystal is governed by two nested lightcones, the theory shows features absent in ordinary (quantized) Maxwell electrodynamics.
  We then compute a QEI bound for worldlines of inertial `subluminal' observers, which generalizes known results from the Maxwell theory.
  Finally, it is shown that the QEIs fail along trajectories that have velocity vectors which are timelike with respect to only one of the lightcones.
\end{abstract}
\bigskip

%******************************************************************************%
\section{Introduction}

The phenomenon of birefringence provides a vivid illustration of the difference between electrodynamics in media and in vacuum -- the propagation of light is governed by two lightcones, neither of which need be that of the background spacetime.
At the theoretical level, neither the constitutive law $H=H(F)$ relating the electromagnetic induction to the field strength, nor the Maxwell equations $\dif H = J$ and $\dif F = 0$, need make reference to any metric structure. Consequently, general electrodynamics can display a much richer causal structure than the vacuum situation in which $H=\mathop{\star} F$, where $\star$ is the Hodge operator induced by the metric.
This pre-metric viewpoint on electromagnetism can be derived from basic principles \cite{Hehl,Lindell} and has been studied both from a phenomenological viewpoint \cite{Hehl20081141, obukhov:2012, Schuller:2009hn} and also for its technical and conceptual interest as an example of a theory based on non-metric structures \cite{drummond, Itin:2015tdi, Punzi:2007di, Rubilar:2007qm}, including the interesting situation where the constitutive law is obtained from a more general geometric structure such as an area metric, which can appear \ia\ as an effective background in quantum electrodynamics on curved backgrounds at first order~\cite{drummond}.

The present paper concerns quantized pre-metric electrodynamics \cite{Rivera:2011rx}, recently formulated in terms of the $1$-form potential by two of us~\cite{Pfeifer:2016har}.
We will investigate properties of its energy density, particularly the extent to which it can assume negative expectation values.
In quantum field theory (QFT) it has long been known that pointwise positivity of the energy density is incompatible with standard assumptions~\cite{EpsGlaJaf:1965}.
Therefore the energy density can exhibit negative expectation values and, at any given point, is typically unbounded from below as a function of the state.
In various theories, however, it turns out that local averages of the energy density are bounded below by Quantum Energy Inequalities (QEIs, also called quantum inequalities).
QEIs have been proved for a variety of free fields in flat and curved spacetimes (see~\cite{ford:ae,Ford:1991,FordRoman:1995,Flanagan:1997,PfenningFord_static:1998,FewsterEveson:1998} for early results and~\cite{Ford:2009vz,Fewster:2012yh,Fewster2017QEIs} for reviews and references) and also for non-free models including a large class of conformal field theories in $2$-dimensions~\cite{FewHoll:2005} and the massive Ising model~\cite{BosCadFew13}.
In~\cite{Fewster:1999gj}, for example, it was shown that smooth local averages of the energy density of a free scalar field along arbitrary smooth timelike curves in any globally hyperbolic spacetime obey QEIs valid in all Hadamard states of the theory (the most general class regarded as physically relevant), and analogous QEIs hold for vacuum electromagnetism~\cite{fewster:2003}.
Our purpose here is to extend these results, for the first time, to a pre-metric theory.

Several aspects of the theory must be reconsidered in the pre-metric setting, because their usual formulation depends on the spacetime metric.
For instance, the energy density is normally defined as a contraction of the stress-energy tensor with a timelike vector, while the QEIs hold along timelike curves, but not along null curves~\cite{FewsterRoman:2003} or over spatial volumes~\cite{FordHelferRoman:2002} or, consequently, along spacelike curves.
Moreover, the defining property of Hadamard states is that their singularity structure is determined by the metric~\cite{KayWald:1991,radzikowski:1996}.
At the outset, therefore, it is not clear how to proceed in a theory with two lightcones, for example, nor is it clear what QEIs can be expected.

To be specific, we consider a general class of electrodynamic theories with spacetime dependent local and linear constitutive laws possessing a single pair of \emph{hyperbolicity cones} in the cotangent bundle (see Sect.~\ref{sub:pmedyn:fresnel} below) of which one can be selected as `positive frequency' (while the other is its exact opposite).
This assumption does not exclude the possibility that there is more than one lightcone, and is compatible with the lightcone structure of a birefringent uniaxial medium, for example.
In this situation one may classify the trajectories on spacetime according to their velocity tangent vectors as subluminal, interluminal or superluminal.
The subluminal trajectories are followed by the \emph{admissible observers} identified in~\cite{Raetzel:2010je} and generalize the notion of timelike curves (travelling more slowly than all light rays) in metric background geometry; by contrast, interluminal observers travel faster than some (but not all) light rays, while superluminal observers travel faster than all light rays.
The classical energy density may be defined along any future-pointing observer trajectory, which in general may be sub- or interluminal (and in some cases even superluminal, see Sect.~\ref{sub:pmedyn:obs}), as a component of a kinematic energy-momentum pseudo $3$-form (replacing the stress-energy tensor). Particular importance will attach to those trajectories along which the (classical) energy density is everywhere non-negative and vanishes only where the field strength does.
We refer to this as the strict weak energy condition (sWEC).

Our main general result is that the quantized energy density obeys a QEI along any future-pointing subluminal observer trajectory for which the classical sWEC holds, in any state obeying a microlocal spectrum condition of the type studied in~\cite{Pfeifer:2016har} and enlarged upon here.
The main problem is to write the energy density in a sum-of-squares form; after that, the argument proceeds more or less as in~\cite{Fewster:1999gj,fewster:2003} taking account of the different form of the microlocal spectrum condition in the present case.
The argument is fully rigorous, making use of microlocal techniques.
As the analysis of~\cite{Pfeifer:2016har} was restricted to translationally invariant constitutive laws, we have to supplement our hypotheses with assumptions that the QFT exists (in the expected form).
In due course it is hoped to address the conditions on the constitutive law under which these assumptions can be proved.

The general QEI is illustrated for the constitutive law corresponding to a translationally invariant uniaxial birefringent medium, in which we are able to compute the finite QEI bound explicitly for subluminal trajectories moving at uniform velocity relative to the medium.
Light propagation is governed by two lightcones, which are nested and touch along a pair of opposing generators.
The outer lightcone in the tangent bundle (corresponding to the inner lightcone in the cotangent bundle) determines the propagation of ordinary (`fast') rays, while the inner lightcone in the tangent bundle governs the extraordinary (`slow') rays.
Subluminal observer trajectories have velocities less than the speed of slow light and the QEI bound is indeed finite for such, but due to the absence of boost and rotational symmetry, which is broken by the preferred direction given by the optic axis, the bound depends on the rapidity with respect to the rest frame of the crystal and the angle to the optical axis of the subluminal observer. The QEI bound diverges as the velocity vector of the trajectory of the observer approaches the inner lightcone.
This leaves open the question of whether there is any constraint on energy densities along faster trajectories, because the QEI bounds are not expected to be sharp, so a divergence in the bound does not imply that this is actually exploited by states of the theory.
We are able to answer this question negatively by explicitly constructing a family of single-particle states whose averaged energy densities may be made arbitrarily negative along trajectories moving at `interluminal' velocities, \ie, between the slow and fast speeds of light in the given direction.
A point of interest here is that the usual counterexamples to the existence of QEIs for spatial or null averaging~\cite{FordHelferRoman:2002,FewsterRoman:2003} are based on superpositions of the vacuum with a two-particle state and involve some careful estimates; here, we are able to give a much more direct and transparent example.
One point that we do not address, however, is whether there might be components of kinematic energy-momentum other than the energy density that have finite QEI bounds along trajectories moving faster than slow light.

We begin with a short review of pre-metric electrodynamics and a thorough extended discussion of the notion of observers in the pre-metric setting in Sect.~\ref{sec:pmedyn}, where we introduce all notions and notations needed in this article and rewrite the electromagnetic energy density of pre-metric electrodynamics into a form well-adapted for quantum energy inequalities.
In the following Sect.~\ref{sec:qei}, we define a (classical) point-split version of the energy density in pre-metric electrodynamics and, after quantizing it, give a general proof of the quantum energy inequality.
We then turn to the explicit example of the uniaxial crystal in Sect.~\ref{sec:uniaxial}, where we derive the two-point function of a ground state for electrodynamics inside the crystal.
In Sect.~\ref{sec:uniaxial-qei} we derive the quantized point-split energy density along subluminal observer trajectories (\ie, slower than the slow speed of light) explicitly.
The resulting quantity is used to compute the QEI bound for these observers.
We then show, by explicit constructions, that the energy density along worldlines with interluminal velocities is not bounded from below, so there are no QEIs for such trajectories.

%******************************************************************************%
\section{Pre-metric electrodynamics}
\label{sec:pmedyn}

Let us recapitulate some basic elements of pre-metric electrodynamics, following~\cite{Hehl}.
In this approach, electrodynamics is formulated quite generally by the equations of motion
\begin{equation}\label{eq:FH-field-eqs}
  \dif A = F,
  \quad
  \dif H = J,
\end{equation}
for the \emph{electromagnetic vector potential} $1$-form $A$, the \emph{field strength} $2$-form $F$, the \emph{induction} pseudo-$2$-form $H$ and the \emph{current} pseudo-$3$-form~$J$.
The physical properties of the electromagnetic medium are encoded in the \emph{constitutive relation}
\begin{equation*}
  H = \conlaw F
\end{equation*}
between $H$ and $F$. There remains a gauge freedom $A\mapsto A+\dif \lambda$ in the
potential $A$. Putting these various equations together one obtains
\begin{equation}\label{eq:A-field-eq}
  \dif \conlaw \dif A = J.
\end{equation}
Once $A$ is obtained (up to the gauge freedom) $F$ and $H$ can be derived from it.
Note that exactness of $F$ follows from the assumption of magnetic flux conservation adoped as an axiom in~\cite{Hehl} and does not require any assumptions on the spacetime topology. (Had one adopted $\dif F=0$ as a starting point, one would need to assume additionally trivial first de Rham cohomology.)

In this paper we adopt the setting of \emph{local and linear pre-metric electrodynamics}, in which the map $\conlaw$ can be expressed using a \emph{constitutive tensor} $\kappa_{ab}{}^{cd}$ so that
\begin{equation}\label{eq:kappa-constitutive}
  (\conlaw F)_{ab} = \frac{1}{2} \kappa_{ab}{}^{cd} F_{cd}.
\end{equation}
Thus we disregard the non-local and non-linear features often exhibited by realistic media.
In general the constitutive tensor is a spacetime dependent pseudo-tensor field.
The methods described in this and the next section are general enough to encompass also spacetime dependent constitutive laws.
Only in Sect.~\ref{sec:uniaxial} do we restrict our considerations to a constitutive law which is constant throughout spacetime.

This is the moment for a brief intermezzo on Levi-Civita symbols, of which we will use two: the first, $\varepsilon^{abcd}$, is the totally antisymmetric rank-$\binom{4}{0}$ pseudo-tensor density of weight $+1$ whose components obey $\varepsilon^{0123}=1$ in every coordinate chart, while the second, $\hat\varepsilon_{abcd}$, is the totally antisymmetric rank-$\binom{0}{4}$ pseudo-tensor density of weight $-1$ with components obeying $\hat\varepsilon_{0123}=1$ in every coordinate chart.
Evidently, the Levi-Civita symbols can attain the numerical values $+1, -1, 0$ in coordinate charts.
Moreover, $\varepsilon^{abcd} \hat\varepsilon_{abcd} = 4!$, but note that the two Levi-Civita symbols cannot be directly identified with each other, thus justifying the notation with and without a hat.
This is in stark contrast with the metric situation, where the Levi-Civita symbols can be transformed into one another (up to a sign depending on the signature of the metric) by raising and lowering indices.

The Levi-Civita symbol allows us to express the constitutive law~\eqref{eq:kappa-constitutive} using the so-called \emph{constitutive density}, a tensor density $\chi^{abcd}$ of weight $+1$ defined so that
\begin{equation}\label{eq:chi-constitutive}
  (\conlaw F)_{ab} = \frac{1}{4} \hat\varepsilon_{abcd} \chi^{cdef} F_{ef}.
\end{equation}
Often it is more convenient to use $\chi$ rather than $\kappa$.
From~\eqref{eq:chi-constitutive}, we immediately read off the antisymmetry in the first and second pair of indices:
\begin{equation*}
  \chi^{abcd} = \chi^{[ab][cd]}.
\end{equation*}
Additionally we assume that~\eqref{eq:FH-field-eqs} can be derived from an action (\ie, it is non-dispersive), which leads to the additional symmetry
\begin{equation*}
  \chi^{[ab][cd]} = \chi^{[cd][ab]}.
\end{equation*}

Finally, using the constitutive density, the field equations~\eqref{eq:A-field-eq} can be rewritten as
\begin{equation}\label{eq:A-field-coords}
  (PA)^a \defn \partial_b ( \chi^{abcd} \partial_c A_d ) = j^a,
\end{equation}
where $j^a = \varepsilon^{abcd} J_{bcd} / 3!$ is the \emph{current density}, obeying the conservation law $\partial_a j^a = 0$.
We emphasize that~\eqref{eq:A-field-coords} is indeed a covariant equation due to the tensor density and antisymmetry properties of $\chi$.
Indeed, $\partial_a$ here is just the covariant derivative with respect to any affine connection.

%++++++++++++++++++++++++++++++++++++++++++++++++++++++++++++++++++++++++++++++%
\subsection{Fresnel polynomial and the quasi-inverse of the principal symbol}
\label{sub:pmedyn:fresnel}

In this section we briefly introduce and define a `quasi-inverse' of the principal symbol of the field equations~\eqref{eq:A-field-coords}.
As described in~\cite{Pfeifer:2016har}, this quasi-inverse can be used to construct the Green functions of the theory if the constitutive law is constant.
In Sect.~\ref{sub:uniaxial:propagators} we will perform this construction for the uniaxial crystal.

The principal symbol of~\eqref{eq:A-field-coords} is
\begin{equation*}
  \mathcal{M}^{ab}(k) = \mathcal{M}^{(ab)}(k) = \chi^{acbd} k_c k_d.
\end{equation*}
We immediately notice that
\begin{equation*}
  \mathcal{M}^{ab}(k) k_a = 0 = \mathcal{M}^{ab}(k) k_b,
\end{equation*}
which reflects the gauge freedom in~\eqref{eq:A-field-eq} as well as the conservation of the current density -- two sides of the same coin.

For each non-zero covector $k$, which may also be complex , choose a vector $\kappa(k)$ such that $k \cdot \kappa(k) = 1$.
It is convenient (and later indeed necessary) to choose $\kappa(k)$ homogeneous of degree~$-1$ in~$k$ for almost all $k$ and henceforth this will be assumed.
The \emph{Fresnel polynomial} is defined (up to an overall sign -- see below) as
\begin{align*}
  \mathcal{G}(k)
  &\defn \adj(\mathcal{M})_{ab}(k) \kappa^a(k) \kappa^b(k) \\
  &\mathrel{\phantom{\defn}\mathllap{=}} \frac{1}{4!} \hat\varepsilon_{c_1 a_1 a_2 a_3} \hat\varepsilon_{d_3 b_1 b_2 b_3} \chi^{a_1 c_1 b_1 d_1} \chi^{a_2 c_2 b_2 d_2} \chi^{a_3 c_3 b_3 d_3} k_{d_1} k_{c_2} k_{d_2} k_{c_3},
\end{align*}
where $\adj(\mathcal{M})$ denotes the adjugate matrix of $\mathcal{M}$.
Clearly it is a density of weight $+1$ and a homogeneous polynomial of order~$4$ in $k$.
Its zeros are the characteristic wave covectors~$k$ which represent light rays in the geometrical optics approximation.
It was first found in~\cite{Rubilar:2007qm} in the study of light propagation in pre-metric linear electrodynamics.
Moreover, the Fresnel polynomial determines whether~\eqref{eq:A-field-coords} possesses a well-posed initial value problem, which it does if $\mathcal{G}$ is a so-called hyperbolic polynomial~\cite{Raetzel:2010je,Pfeifer:2016har}.

We say that the Fresnel polynomial is \emph{hyperbolic} at $x \in M$ with respect to a covector $n$ if $\mathcal{G}(x, n)\neq 0$ and
$t \mapsto \mathcal{G}(x, \xi + t n)$ has only real roots for real covectors $\xi$.
The covectors $n$ for which $\mathcal{G}(x)$ is hyperbolic at the spacetime point $x$ form open convex cones $\Gamma_x(n) \subset T_x^{\mathrlap{*}}M$, called \emph{hyperbolicity cones}.
It can be shown that hyperbolicity cones always exist in pairs $\Gamma_x(n)$ and $\Gamma_x(-n) = - \Gamma_x(n)$, \ie, if $\mathcal{G}(x)$ is hyperbolic with respect to~$n$, it is also hyperbolic with respect to~$-n$.
If it is possible to choose a smooth distribution $\Gamma = \bigsqcup_{x\in M}\Gamma_x$ of hyperbolicity cones for~$\mathcal{G}$, we say that the Fresnel polynomial is hyperbolic on $M$ with respect to $\Gamma$.
The hyperbolicity double-cones $\Gamma \cup (-\Gamma)$ are the generalizations of the cones of past and future pointing timelike covectors from Lorentzian geometry.
Given such a choice we call the selected covectors in $\Gamma$ subluminal future-pointing covectors.
As we will see in the next section, they can be used to identify future-pointing vectors, \ie, directions, on spacetime.
In metric geometry one classifies subluminal covectors according to the sign of their Lorentzian `norm'.
Depending on the signature convention for the metric, this can be either positive or negative.
Similarly the sign of $\mathcal{G}$ is constant on any hyperbolicity cone; in this article we choose it (without loss) to be positive inside~$\Gamma$.

Thus a hyperbolic Fresnel polynomial defines the causal structure of the theory, which is usually determined by the Lorentzian spacetime metric in ordinary Maxwell vacuum electrodynamics.
It determines the timelike and null covectors relevant in the theory.
Further details on hyperbolicity cones are discussed in~\cite{Hoermander2} and~\cite{Raetzel:2010je,Pfeifer:2016har}.
We will always assume that the Fresnel polynomial is hyperbolic on spacetime.
This condition is comparable to the condition that the spacetime metric is Lorentzian and does not degenerate at any point of spacetime.

With help of the gauge fixing vector field and the Fresnel polynomial we can construct a pointwise `quasi-inverse' $\mathcal{E}$ of the principal symbol $\mathcal{M}$; see~\cite{Pfeifer:2016har} for details.
It is given by
\begin{equation}
	\mathcal{E}_{ab}(k) \defn \frac{\mathcal{Q}_{cd}(k) \pi^c_a(k) \pi^d_b(k)}{\mathcal{G}(k)},
\end{equation}
where $\pi^c_a(k) = \delta^c_a - \kappa^c(k) k_a$ are projectors onto a subspace $V_k$ of $T_x^{\mathrlap{*}}M$ complementary to the ray of $k$, and $\mathcal{Q}$ is determined by the second adjugate of the principal symbol
\begin{align}
  \mathcal{Q}_{ab}(k)
  &\defn \mathrm{adj}_2(\mathcal{M})_{abcd}(k) \kappa^c(k) \kappa^d(k) \\
  &\mathrel{\phantom{\defn}\mathllap{=}} \frac{1}{8} \hat\varepsilon_{b c_1 a_1 a_2} \hat\varepsilon_{a d_2 b_1 b_2} \chi^{a_1 c_1 b_1 d_1} \chi^{a_2 c_2 b_2 d_2} k_{d_1} k_{c_2}.
\end{align}
Although $\mathcal{G}$ and $\mathcal{Q}$ are gauge independent, it is evident that $\mathcal{E}$ depends on the choice of the gauge fixing vector field~$\kappa$.

The quasi-inverse satisfies $\mathcal{M}^{ca}(k) \mathcal{E}_{ab}(k) = \pi^c_b(k) = \mathcal{E}_{ba}(k)\mathcal{M}^{ac}(k)$ whenever $\mathcal{G}(k) \neq 0$.
It is a true inverse of $\mathcal{M}$, regarding the latter as a map from
$V_k\subset T_x^{\mathrlap{*}}M$ to the annihilator of~$k$ in~$T_xM$.
Note also that $\mathcal{E}_{ab}(k)$ is homogeneous of degree~$-2$ in~$k$ and thus, in particular, $\mathcal{E}_{ab}(k) = \mathcal{E}_{ab}(-k)$.

%++++++++++++++++++++++++++++++++++++++++++++++++++++++++++++++++++++++++++++++%
\subsection{Observers}
\label{sub:pmedyn:obs}

As there is no spacetime metric in pre-metric electrodynamics, the description of legitimate observer trajectories requires additional discussion.
In fact, the motion of observers need have no relation to the laws governing propagation of light, as is the case in the phenomenon of Cherenkov radiation.
Nonetheless, in our discussion of the QEIs, it will be necessary to classify observer worldlines according to whether their tangent vectors are slower than all light (which we call \emph{subluminal}), faster than some but not all light (\emph{interluminal}), or faster than all light (\emph{superluminal}).
Standard Maxwell electrodynamics, governed by a single lightcone, excludes the possibility of interluminal vectors.

Our classification will rely on methods and results developed in~\cite{Raetzel:2010je} and requires some additional technical assumptions on the Fresnel polynomial $\mathcal{G}$: specifically, we will assume that $\mathcal{G}$ is \emph{reduced}, \emph{bihyperbolic},
\emph{energy-distinguishing} and \emph{time-distinguishing}, all of which will be explained below.
(These conditions have been identified as important to obtain a reasonable physical theory~\cite{Raetzel:2010je}; however, it has not been determined whether they are all independent, or whether bihyperbolicity might also imply the time- and energy-distinguishing properties.)

Under these conditions, the null covectors split into positive and negative frequency cones, defined consistently in terms of the sign of their contraction with subluminal vectors.
Having identified these cones we find that, conversely, subluminal vectors can be alternatively characterized among future-directed vectors as the connected component of vectors which has positive contractions with all positive frequency null covectors.

This characterization will play an important role in the derivation of the QEI, which will hold for the subluminal observers.
Later, in Sect.~\ref{sub:uniaxial-qei:interluminal}, it will be seen that energy densities observed by interluminal observers do not obey (state independent) QEIs.

On a point of notation, we will need to define a number of subsets $U_x$ of the tangent and cotangent spaces $T_xM$ and $T_x^{\mathrlap{*}}M$.
In such cases, the same symbol without the subscript will denote the corresponding subset $U = \bigsqcup_{x\in M}U_x$ of the bundles $TM$ or $T^*\!M$.

Our starting point is the Fresnel polynomial $\mathcal{G}$, and a choice of hyperbolicity cone $\Gamma$, on which $\mathcal{G}$ is positive.
In addition, we can already identify the set of null covectors
\begin{equation*}
  \mathcal{N}_x \defn \big\{ k \in T_x^{\mathrlap{*}}M \setminus \{0\} \;\big|\; \mathcal{G}(x, k) = 0 \big\},
\end{equation*}
which governs the propagation of massless momenta in geometric optics.
In metric geometry the set $\mathcal{N}_x$ bounds the hyperbolicity cone, $\mathcal{N}_x=(\partial \Gamma)\setminus\{0\}$, but in general it can be larger, as the example of the birefringent crystal nicely demonstrates.
Given these ingredients, we may pick out cones of future and past-pointing tangent vectors by
\begin{equation*}
  \Gamma_x^\pm \defn \bigl\{z\in T_xM\ \big|\ \pm k \cdot z > 0 \text{ for all } k \in \Gamma_x \bigr\}.
\end{equation*}
In the special case $M=\RR^4$, and identifying $M$ with $T_0M$, the closure of $\Gamma_0^+$ contains the support of a fundamental solution of the constant-coefficient partial differential operator $\mathcal{G}(i\partial)$ \cite[Thm.~12.5.1]{Hoermander2}.

In metric geometry, $\Gamma_x^+$ would be the forward causal cone at $x$ and observers with tangent vectors in $\Gamma^+$ could agree on a partition of $\mathcal{N}$ into positive and negative frequency cones.
This is not possible in general.
To make progress, we must study the propagation of massless particles in the geometric optics limit, described in a phase space picture by the Helmholtz action
\begin{equation*}
  S[x,k,\lambda] = \int \bigl( k \cdot \dot{x} - \lambda \mathcal{G}(x,k) \bigr)\, \dif\tau,
\end{equation*}
where $\tau\mapsto (x(\tau),k(\tau))$ is any parameterisation of the trajectory in $T^*\!M$ and $\lambda$ is a Lagrange multiplier, restricting to null covectors.
One passes to configuration space by eliminating $\lambda$ and $k$ as functions of $x$ and $\dot x$ with help of the equations of motion obtained by varying the Helmholtz action.
The result is a new action given by
\begin{equation*}
  S[x,\mu] = \int \mu\, \mathcal{G}^\#(x, \dot x)\, \dif\tau,
\end{equation*}
where $\mu$ is a new Lagrange multiplier function and $\mathcal{G}^\#(x,\dot{x})$ is the so-called \emph{dual polynomial} on~$TM$, which is determined up to an irrelevant overall factor by the above procedure.
The Lagrange multiplier $\mu$ implies that the corresponding tangent vectors
$z$ to solution curves lie in the set of lightlike vectors
\begin{equation*}
  \mathcal{N}_x^{\#} \defn \big\{  z \in T_xM \setminus \{0\} \;\big|\; \mathcal{G}^\#(x, z) = 0 \big\}.
\end{equation*}

At this point, we introduce two of our assumptions on $\mathcal{G}$.
First, we assume that the lightlike tangent vectors $\mathcal{N}^\#$ can be partitioned into (necessarily disjoint) future- and past-pointing cones
\begin{equation*}
  (\mathcal{N}^{\#}){}^\pm = \mathcal{N}_x^{\#} \cap \Gamma^\pm.
\end{equation*}
In this situation, where  $\mathcal{N}^{\#}=(\mathcal{N}^{\#}){}^+ \cup (\mathcal{N}^{\#}){}^-$, we say that $\mathcal{G}$ is \emph{time-distinguishing}.
Second, we will assume that $\mathcal{G}$ is \emph{bihyperbolic}, which means that both $\mathcal{G}$ and the dual polynomial $\mathcal{G}^\#$ are hyperbolic on $M$.
Let $\Gamma^\#\subset TM$ be a hyperbolicity cone for $\mathcal{G}^\#$, chosen to be future-directed, \ie, $\Gamma^\#\subset\text{int}(\Gamma^+)$.\footnote{Here, we adapt an argument from~\cite[p.12]{Raetzel:2010je}, switching the roles of $\mathcal{G}$ and $\mathcal{G}^\#$, to show that a bihyperbolic and time distinguishing Fresnel polynomial always has a hyperbolicity cone for $\mathcal{G}^\#$ contained in $\Gamma^+$, and in fact within the interior thereof, given that hyperbolicity cones are open.} It may be assumed without loss of generality that $\mathcal{G}^\#$ is positive on $\Gamma^\#$.

We now come to the classification of non-lightlike tangent vectors at each point $x\in M$, based on the connected component of $T_xM\setminus (\mathcal{N}_x^\#\cup\{0\})$ to which they belong.
Namely,
\begin{itemize}
  \item $\Gamma^\#_x$ is the component of future-directed \emph{subluminal} vectors;
  \item any component, other than $\Gamma^\#_x$, whose boundary is contained in $(\mathcal{N}_x^{\#}){}^+\cup\{0\}$ consists of future-directed \emph{interluminal} vectors;
  \item any component whose boundary meets both $(\mathcal{N}_x^{\#}){}^+$ and $(\mathcal{N}_x^{\#}){}^-$ consists of \emph{superluminal} vectors;
  \item $z\in T_xM$ is past-directed subluminal (resp., interluminal) if $-z$ is future-directed subluminal (resp., interluminal);
\end{itemize}

\begin{figure}
  \centering
  \begin{minipage}{.45\textwidth} \centering
    \includegraphics[height=4cm]{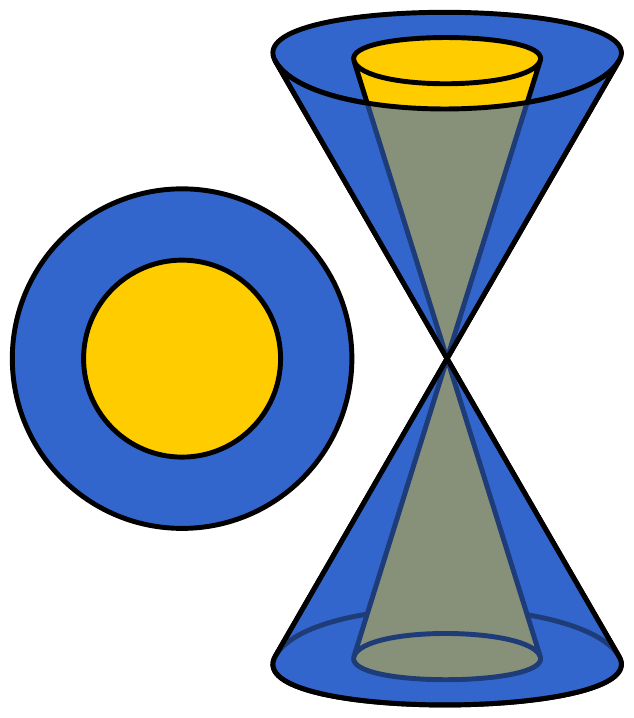} \\
    (a)
  \end{minipage}
  \begin{minipage}{.45\textwidth} \centering
    \includegraphics[height=4cm]{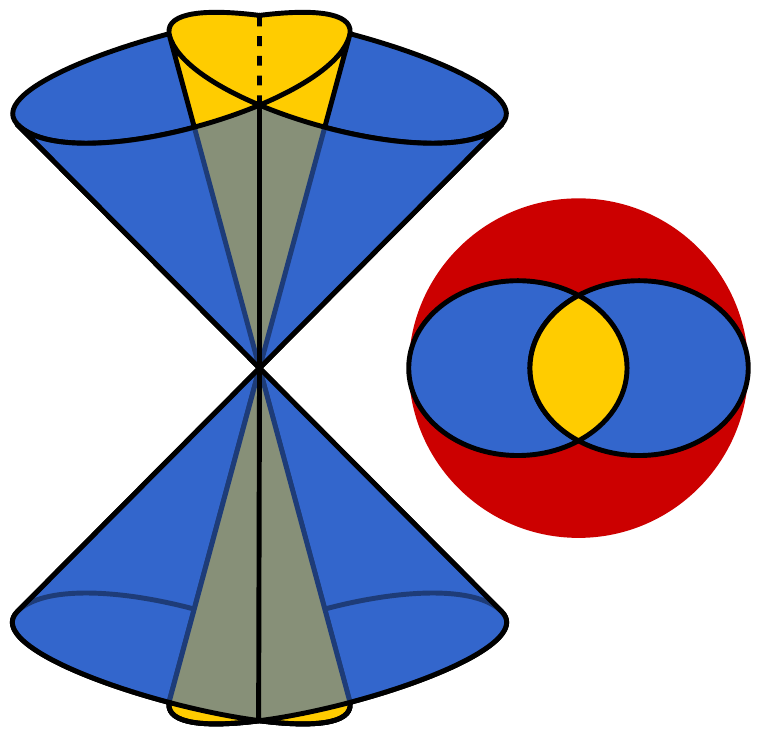} \\
    (b)
  \end{minipage}
  \caption{Illustration of the classification of tangent vectors for two different bi-metric dual polynomials $G^\#$, showing nesting (a) or crossing (b) of light cones. The circular pictures are slices through the future half of the cones. Vectors are classified as subluminal (yellow), interluminal (blue) and null (black). All other vectors are superluminal. In (b) the bounded disc of future-directed superluminal vectors is coloured red.}
  \label{fig:observers}
\end{figure}

A few remarks are appropriate here, which we illustrate with the sketch in Fig.~\ref{fig:observers}.
First, subluminal vectors were identified in~\cite{Raetzel:2010je} as the tangent vectors to worldlines of `admissible observers'; here, we prefer to use the term `subluminal', the yellow sets in the illustration, and to interpret the boundary $(\partial\Gamma^\#)\setminus\{0\}\subset (\mathcal{N}^{\#}){}^+$ as the cone of slowest future-pointing lightlike vectors.
The designation of `interluminal' vectors as those trapped between either future or past pointing lightlike sets should be clear, they are blue in the picture.
In metric geometry, there are no future or past pointing superluminal vectors. Here, they may exist on the basis of the rich null-structure available in pre-metric electrodynamics, see the red set in the right part of the sketch.

Observer worldlines can now be identified as future-directed spacetime trajectories $\gamma$, \ie, their tangent satisfies $\dot{\gamma}\in \Gamma^+$.
They can be labeled as being sub-, inter- or superluminal depending on whether $\dot{\gamma}$ is everywhere sub-, inter- or superluminal according to the definition given above.
In the standard Maxwell electrodynamics only subluminal observers exist, while in the example of bi-metric Fresnel polynomials with nested lightcones sub- and interluminal observers come into play, see the left part of Fig.~\ref{fig:observers}, Fig.~\ref{fig:bimetric} and the discussion of the uniaxial crystal in Sect.~\ref{sec:uniaxial}.
Bi-metric light propagation with overlapping lightcones, as in the right part of Fig.~\ref{fig:observers}, is an example for the existence of sub-, inter- and superluminal observer directions (see also ~\cite[Fig.~7]{Raetzel:2010je}).

We continue to develop the physical interpretation of the subluminal vectors, aiming for an alternative characterization for use in Sect.~\ref{sub:qei:proof}.
Just as the hyperbolicity cone $\Gamma\subset T^*\!M$ allowed the definition of cones of future- and past-directed vectors, the hyperbolicity cone $\Gamma^\#\subset TM$ determines cones $\Gamma^{\#\pm}$ of positive and negative frequency covectors by
\begin{equation*}
  \Gamma_x^{\#\pm} \defn \bigl\{k\in T_x^{\mathrlap{*}}M\ \big|\ \pm k \cdot z > 0 \text{ for all } z \in \Gamma^\#_x \bigr\}.
\end{equation*}
A Fresnel polynomial $\mathcal{G}$ is said to be \emph{energy-distinguishing} if and only if every null covector has either positive or negative frequency, \ie,  $\mathcal{N} = \mathcal{N}^+ \cup \mathcal{N}^-$, where\footnote{Note that in the previous article~\cite{Pfeifer:2016har} the definition of the set $\mathcal{N}^\pm$ was incomplete in the sense that the sets defined there did not contain all future/past-pointing null directions for general Fresnel polynomials.}
\begin{equation}\label{eq:Npm}
  \mathcal{N}_x^{\pm} \defn \mathcal{N}_x \cap \Gamma_x^{\#\pm} = \big\{ k \in \mathcal{N}_x \;\big|\; \pm k \cdot z > 0 \text{ for all } z \in \Gamma_x^\# \big\}.
\end{equation}

It can now be proven that for bihyperbolic and time- and energy-distinguishing Fresnel polynomials which are also \emph{reduced} -- that is, in any factorisation of $\mathcal{G}$ into polynomials with real coefficients there are no repeated non-constant factors -- the subluminal vectors can be alternatively characterized among future-pointing vectors as those having non-negative contractions with all positive frequency null covectors.
In other words, for $z\in \text{int}(\Gamma^+_x)$,
\begin{equation}\label{eq:timlike-subluminal}
	z \in \Gamma^\#_x \Longleftrightarrow k \cdot z > 0 \textrm{ for all } k \in \mathcal{N}^+_x.
\end{equation}
Physically, this means that the subluminal observers are precisely those that agree on $\mathcal{N}^+$ as indeed having positive frequency.

To establish~\eqref{eq:Npm}, consider first any subluminal $z \in \Gamma^\#_x\subset \text{int}(\Gamma^+_x)$; by definition of $\mathcal{N}^+_x$ it holds that $k \cdot z > 0$ for all $k \in \mathcal{N}^+_x$.
To establish the reverse implication we employ the invertible Legendre map $L_x$ which maps $\Gamma_x$ into $T_xM$
\begin{equation}\label{eq:Legendre}
  L_x : \Gamma_x \rightarrow T_xM,\quad k \mapsto \frac{1}{4\mathcal{G}(x,k)}\frac{\partial\mathcal{G}(x,k)}{\partial k_a}.
\end{equation}
As shown in the third and fourth lemmas in~\cite[Sect.~VII]{Raetzel:2010je}, the range of $L_x$ is $L_x(\Gamma_x)=\text{int}(\Gamma_x^+)$ and contains $\Gamma^\#_x$.
As a map $L_x:\Gamma_x \rightarrow \text{int}(\Gamma_x^+)$ an inverse of the Legendre map exists $L^{-1}_x:\text{int}(\Gamma_x^+)\rightarrow \Gamma_x$.
It has the property that the inverse image of $\Gamma_x^\# \subset \text{int}(\Gamma_x^+)$,
\begin{equation*}
  S_x \defn L_x^{-1}(\Gamma_x^\#) \subset \Gamma_x,
\end{equation*}
is precisely the cone of \emph{stable momenta}: those massive momenta $k\in\Gamma_x$ that cannot lose energy by emitting Cherenkov radiation while remaining on the same mass-shell (level sets of $\mathcal{G}(x,\cdot{})$ within $\Gamma_x$).
Now if $z\in \text{int}(\Gamma_x^+)$ but $z\notin \Gamma^\#_x$, then $L_x^{-1}(z)\notin S_x$.
This implies in turn, see~\cite[Sect.~X]{Raetzel:2010je}\footnote{Observe that there is a sign error in~\cite[Sect.~X]{Raetzel:2010je} in the statement of conditions under which a massive momentum $p$ may radiate positive frequency massless momentum in a Cherenkov-like process.
The condition stated there is that there must exist a massless momentum $r \in \mathcal{N}_x^+$ such that $r \cdot L_x(p) > 0$.
However, doing the calculation with the conventions used in~\cite[Sect.~X]{Raetzel:2010je} actually yields the opposite condition, namely $r \cdot L_x(p) < 0$.}, that there exists a null covector $k \in \mathcal{N}^+_x$ such that $0 > k \cdot L_x(L_x^{-1}(z)) = k \cdot z$.
Taking the contrapositive, the proof of~\eqref{eq:timlike-subluminal} is complete.

This proof also enables us to characterize the positive and negative frequency null covectors as
\begin{equation*}\label{eq:Npm2}
  \mathcal{N}_x^{\pm} = \big\{ k \in \mathcal{N}_x \;\big|\;  \pm k_a \partial\mathcal{G}(x, n)/\partial n_a > 0 \text{ for all } n \in S_x \big\}.
\end{equation*}
This is true since $\frac14 k_a \partial\mathcal{G}(x,n)/\partial n_a = \mathcal{G}(x,n) k \cdot L_x(n) > 0$ for $n \in S$ since $L_x(n) \in \Gamma_x^\#$ and we have assumed that $\mathcal{G}(x,n) > 0$ for all $n\in \Gamma_x$ so that $\mathcal{G}(x,n)\, k \cdot L_x(n) > 0$ is equivalent to $k \cdot L_x(n) > 0$.
When we prepare for the proof of the QEI in Sect.~\ref{sub:qei:quant_pt_split} the characterizations~\eqref{eq:timlike-subluminal} and~\eqref{eq:Npm2} will ensure that the quantized point-split energy density can be pulled back to subluminal observer trajectories as a bi-distribution.

Summarizing, we have given conditions on the Fresnel polynomial that allow for the classification of tangent vectors as sub-, inter- or superluminal, and provide a corresponding definitions for observers.
Further, the subluminal vectors and positive frequency null covectors have been given alternative characterizations that will be needed in the sequel.
We would like to stress again that only subluminal observers are stable; physical inter- and superluminal observers would emit Cherenkov radiation until they become subluminal.

%++++++++++++++++++++++++++++++++++++++++++++++++++++++++++++++++++++++++++++++%
\subsection{Energy density}
\label{sub:pmedyn:edensity}

In a pre-metric theory, the stress-energy tensor clearly cannot be obtained by variations of an action with respect to the (absent) metric.
It is, however, possible to define the stress-energy tensor of the electromagnetic field on kinematic grounds, see Chaps.~B.2 and~B.5 of~\cite{Hehl} and Sect.~2.8 of~\cite{Pfeifer:2016har} which we follow here, writing the resulting energy density in a form which is suitable for the derivation of the quantum energy inequality.

In pre-metric electrodynamics, the \emph{kinematic energy-momentum} is a pseudo-$3$-form defined in terms of a vector field $N$ by
\begin{equation*}
	T_{N} \defn \frac12 \bigl( F \wedge (\iprod{N} H ) - H \wedge (\iprod{N} F) \bigr).
\end{equation*}
This is motivated physically \cite{Hehl} by the requirement that $\dif T_N$ is related to a component of the Lorentz force by $\dif T_N = (\iprod{N} F) \wedge J$, in the case where $N$ is a symmetry vector field, \viz, if the Lie derivative of the constitutive tensor with respect to the vector field vanishes: $\mathcal{L}_N \kappa_{ab}{}^{cd} = 0$.

The kinematic energy-momentum can be used to define the \emph{energy density} of the electromagnetic field along a worldline $\gamma$ in the following way.
Choose any basis $e=\{e_a\}_{a=0}^3$ of the tangent spaces along $\gamma$ that can be extended smoothly to a contractible neighbourhood $\mathcal{T}$ of $\gamma$ so that $e_0$ coincides with the observer's velocity vector on~$\gamma$, $e_0|_{\gamma(\tau)}=\dot{\gamma}(\tau)$, and is everywhere future-pointing in~$\mathcal{T}$ (\ie, $e_0|_x\in \Gamma^+_x$).
The integral curves of~$e_0$ define a congruence of observer worldlines in~$\mathcal{T}$, and at each point~$x$, the basis~$e$ specifies a system of rods and clocks for the observer at~$x$. Denote the dual basis by $\{e^{*a}\}_{a=0}^3$ and write $u=e_0$, $n=e^{*0}$.
Then the energy density of the electromagnetic field in~$\mathcal{T}$ with respect to the frame~$e$ is
\begin{align}
	\rho = (n\wedge T_{u})(e_0, e_1, e_2, e_3),
\end{align}
which, at any given point $x\in\mathcal{T}$, is the component of the $4$-form $n\wedge T_{u}$ along the observer worldline through $x$ and taken with respect to the observer's frame $e$.
In particular, the energy density along $\gamma$ is obtained by restriction; clearly, it depends both on $\gamma$ and the choice of frame $e$.

The energy density may also be expressed as
\begin{equation}
  \rho = \frac18 \varepsilon(e)^{-1} \chi^{abcd} (F_{ab} F_{cd} - 4 n_a u^e F_{eb} F_{cd}),
\end{equation}
using the constitutive law $H = \conlaw F$, where we have written $\varepsilon(e)=\varepsilon(e_{0},e_{1},e_{2},e_{3})$ for short.
If the frame is obtained from a system of local coordinates, $e_a=\partial/\partial x^a$, the density factor becomes $\varepsilon(e) = 1$ (in those coordinates).

In order to derive the QEI it is useful to give a novel form for $\rho$, which we have not seen discussed elsewhere for pre-metric electrodynamics.
Setting
\begin{equation*}
  \lambda_a^b = \delta_a^b - n_a u^b,
\end{equation*}
a straightforward calculation then shows that
\begin{align}
  \rho &= \frac18 \varepsilon(e)^{-1}\chi^{abcd} \lambda_b^f \lambda_d^{h\vphantom{f}} (\lambda_{a\vphantom{b}}^{e\vphantom{f}} \lambda_{c\vphantom{b}}^{g\vphantom{f}} - 4 n_a u^e n_c u^g) F_{ef} F_{gh} \nonumber \\
  &= \frac18 (\chi^{abcd}_1 + \chi^{abcd}_2) F_{ab} F_{cd}, \label{eq:chi_12-decomp}
\end{align}
where
\begin{align*}
  \chi^{efgh}_1 &\defn \varepsilon(e)^{-1} \chi^{abcd} \lambda_{a\vphantom{b}}^{e\vphantom{f}} \lambda_b^f \lambda_{c\vphantom{b}}^{g\vphantom{f}} \lambda_d^{h\vphantom{f}}, \\
  \chi^{efgh}_2 &\defn -4 \varepsilon(e)^{-1}\chi^{abcd} n_a u^e \lambda_b^f n_c u^g \lambda_d^{h\vphantom{f}}.
\end{align*}
These expressions will allow us to find a point-split energy density in the next section.
They also give some insight into the (classical) positivity of the energy density: $\rho$ is certainly non-negative if both $\chi_1$ and $\chi_2$ determine positive (semi-)definite metrics on the space of $2$-forms.
In fact, this is a necessary and sufficient condition.
To see why, note that $\lambda_{[a}^{c} \lambda_{b]}^d$ projects (at each point $x\in\mathcal{T}$) onto the $3$-dimensional subspace of `magnetic' $2$-forms $u^\perp\wedge u^\perp$, where $u^\perp$ consists of covectors annihilating $u$, while $2n^{\vphantom{d}}_{\smash{[}a} u^c \lambda_{b\smash{]}}^d$ projects onto a complementary $3$-dimensional `electric' subspace $n\wedge u^\perp$.
If $\chi_1^{abcd}G_{ab}G_{cd}<0$, then by defining $F_{ab} = \lambda_{a\vphantom{b}}^{c\vphantom{d}} \lambda_b^d G_{cd}$, one has $\rho=\frac{1}{8}\chi_1^{abcd}G_{ab}G_{cd}<0$.
A similar argument shows that if $\chi_2$ fails to be positive semi-definite, then so does $\rho$.

Thus the \emph{weak energy condition (WEC)}, \ie, non-negativity of $\rho$, is equivalent to positive semi-definiteness of $\chi_1$ and $\chi_2$.

A stronger statement can be made, and will be useful to us in what follows. Suppose $\rho$ is not only non-negative, but vanishes precisely at points of vanishing field strength~$F$:
\begin{equation*}\tag{sWEC}
  \emph{$\rho(x)\ge 0$ for all $x\in\mathcal{T}'$, and $\rho(x)=0$ for some $x\in\mathcal{T}'$ if and only if $F|_{x}=0$}
\end{equation*}
where $\mathcal{T}'$ is a subset of $\mathcal{T}$.
Then we will say that \emph{the strict form of the weak energy condition (sWEC)} holds on $\mathcal{T}'$ with respect to the frame $e$.
The sWEC is equivalent to $\chi_1$ and $\chi_2$ being positive definite on the magnetic and electric subspaces respectively at all points in $\mathcal{T}'$.
For instance, if $\chi_1$ is not positive definite on the magnetic subspace, then there is a non-zero $F_{ab}$ in this subspace for which $\rho=\frac{1}{8}\chi_1^{abcd}F_{ab}F_{cd}\le 0$ and sWEC fails; a similar argument holds in the electric case.
Conversely, if $\chi_1$ and $\chi_2$ are positive definite on their respective subspaces then it is easily seen from the definitions that sWEC holds.

%++++++++++++++++++++++++++++++++++++++++++++++++++++++++++++++++++++++++++++++%
\subsection{Quantization}
\label{sub:pmedyn:quant}

We briefly discuss the quantization of the electromagnetic potential in pre-metric electrodynamics as described in~\cite{Pfeifer:2016har} with the following differences.
Instead of closed $3$-forms we base the quantization here analogously on conserved vector densities with which they are in one-to-one correspondence.
Moreover we use a different sign convention for the Fourier transform $\what{f}(\xi) \defn \int f(x)\, \e^{\im \xi \cdot x}\, \dif^n x$, where $\,\cdot\,$ denotes the Euclidean dot product.

Throughout this section, denote by $j, j'$ arbitrary conserved compactly supported vector densities, and by~$A$ compactly supported $1$-forms.
Further, recall that $(PA)^a=\partial_b (\chi^{abcd} \partial_c A_d)$.
Distributions of some tensorial type are continuous linear functionals on compactly supported densities of the dual tensorial type, \eg, a covector distribution acts on vector densities, and this convention extends in an obvious way to bi-distributions.

In order to formulate the quantum theory, we assume that there is an antisymmetric covector bi\hyp{}distribution~$\sigma$ which restricts to the space of conserved compactly supported vector densities as an anti-symmetric, bilinear form with the property that $\sigma(j, j') = 0$ for all $j'$ if and only if $j = P A$ for some compactly supported $1$-form $A$. If the constitutive law is constant, it was shown in~\cite[\S II.G]{Pfeifer:2016har} that such an anti-symmetric distribution exists and is given by the Pauli--Jordan propagator, \viz, the difference of advanced and retarded Green functions obtained with respect to a choice of gauge. Thus our assumption amounts to a requirement on the global well-posedness and solvability (up to gauge transformations) of the field equation~\eqref{eq:A-field-eq}.
In Sect.~\ref{sec:uniaxial}, we explicitly construct the Pauli--Jordan propagator for the Fresnel polynomial of an uniaxial crystal.

Once $\sigma$ is fixed, the quantization may be performed using well-established methods from algebraic quantum field theory as used for example in mathematical approaches to quantum field theory in curved spacetimes.
Namely, we construct an algebra of quantum fields $\mathfrak{A}$, which is the unital ${}^*$-algebra finitely generated by a \emph{smeared quantum field observables} $\what{A}(j)$ labelled by (complex-valued) compactly supported, conserved vector densities $j$, and satisfying the relations
\begin{description}[labelwidth=8em,align=right]
  \item[Linearity] $\what{A}(\alpha j + \beta j') = \alpha \what{A}(j) + \beta \what{A}(j')$ for all $\alpha, \beta \in \CC$,
  \item[Hermicity] $\what{A}(j)^* = \what{A}(\conj{\jmath})$,
  \item[Field equation] $\what{A}(P A)=0$,
  \item[CCR] $\displaystyle \bigl[\what{A}(j), \what{A}(j')\bigr] = \im \sigma(j, j') \one$;
\end{description}
here, we denote the unit element of~$\mathfrak{A}$ by~$\one$ and make use of our standing conventions on~$j$'s and~$A$'s.

The algebra element $\what{A}(j)$ can be interpreted as a smeared field $\int \!\what{A}_a j^a$ (recall that~$j$ is a vector density of weight~$1$, so no volume element appears); later, we will discuss Hilbert space representations in which this can be taken literally, with $\what{A}_a$ understood as an operator-valued distribution.

It is convenient to identify elements of~$\mathfrak{A}$ corresponding to smeared field strengths: for any smooth compactly supported second rank contravariant tensor density $t$, we define
\begin{equation}\label{eq:smeared_F}
  \what{F}(t) \defn 2\what{A}(\div t),
\end{equation}
where $(\div t)^a = \partial_b t^{[ab]}$ is clearly a conserved vector density; $\what{F}(t)$ can be interpreted as a smeared field $\int \!\what{F}_{ab} t^{ab}$.

The normalized positive functionals on $\mathfrak{A}$ are called \emph{(quantum) states}.
That means, $\Lambda$ is a state on the field algebra $\mathfrak{A}$ if
\begin{description}[labelwidth=8em,align=right]
  \item[Normalization] $\Lambda(\one) = 1\;$ and
  \item[Positivity] $\Lambda(a^* a) \geq 0\;$ for all $a \in \mathfrak{A}$.
\end{description}
Each state $\Lambda$ can be represented by a hierarchy of \emph{$n$-point functions} $(\Lambda_n)_{n \geq 0}$ by setting
\begin{equation*}
  \Lambda_n(j_1, \dotsc, j_n) \defn \Lambda\big(\what{A}(j_1) \,\dotsm\, \what{A}(j_n)\big)
\end{equation*}
for conserved compactly supported vector densities $j_1,\ldots,j_n$, and then extending arbitrarily to general compactly supported vector densities.
In this way the state fixes the $n$-point functions only up to gauge equivalence.

Of particular importance are \emph{quasi-free states} (also called \emph{Gaussian states}).
These states are completely characterized by their two-point function so that all even $n$-point functions are given by sums of products of two-point functions according to a Wick expansion and all odd $n$-point functions vanish.
A two-point function $\Lambda_2$ necessarily satisfies the following relations (recall our standing conventions concerning the symbols $j$ and $A$):
\begin{description}[labelwidth=8em,align=right]
  \item[Positivity] $\Lambda_2(\conj{j}, j) \geq 0$,
  \item[Hermiticity] $\overline{\Lambda_2(j,j')}=\Lambda_2(\conj{j'}, \conj{j})$
  \item[Field equation] $\Lambda_2(j, P A) = 0 = \Lambda_2(P A, j)$,
  \item[CCR] $\displaystyle \Lambda_2(j, j') - \Lambda_2(j', j) = \im \sigma(j, j')$.
\end{description}
In the framework developed in~\cite{Pfeifer:2016har}, physical states in pre-metric electrodynamics are required to obey the \emph{microlocal spectrum condition} ($\mu$SC), a generalization of the Hadamard condition used for QFT in curved spacetimes~\cite{KayWald:1991,radzikowski:1996}:
\begin{description}[labelwidth=8em,align=right]
\item[$\mu$SC] among the gauge equivalent two-point functions $\Lambda_2$ induced by the state $\Lambda$, there should be at least one that is a covector bi-distribution, with wave-front set obeying
\begin{equation}\label{eq:microlocal}
	\WF(\Lambda_2) \subset \mathcal{N}^+ \times \mathcal{N}^- \subset T^*\!M \times T^*\!M
\end{equation}
with $\mathcal{N}^\pm$ as defined in~\eqref{eq:Npm} or equivalently~\eqref{eq:Npm2},
and whose anti-symmetric part is fixed up to smooth terms by the \emph{generalized CCR}\footnote{This condition was implicitly assumed in~\cite{Pfeifer:2016har}; here we make it explicit.}
\begin{equation*}
  \Lambda_2 - \Lambda_2^T = \im\sigma \pmod{C^\infty},
\end{equation*}
where the transposed distribution is defined by $\Lambda_2^T(f,f')=\Lambda_2(f',f)$ for general compactly supported vector densities $f,f'$.
\end{description}

The wave-front set~\cite{Hoermander1} encodes details about the singular structure of a distribution in both configuration and momemtum space. The condition~\eqref{eq:microlocal} asserts that the wave-front set of $\Lambda_2$ consists of pairs $((x_1,k_1),(x_2,-k_2))\in T^*\!M \times T^*\!M$ such that $(x_i,k_i)$ are zeros of the Fresnel polynomial $\mathcal{G}(x_i,k_i)=0$ and $(x_1,k_1)$ lies on the positive frequency null-structure while $(x_2,-k_2)$ lies on the negative frequency null-structure. It is possible to be rather more specific, because the propagation of singularities imposes further relations on the wave-front set. In scalar QFT on curved spacetimes, for example, the pairs $(x_1,k_1)$ and $(x_2,-k_2)$ must be connected by the Hamiltonian flow induced by the principal symbol of the Klein--Gordon equation. We avoid further specification here, partly because it will be unnecessary for our purposes, but also because in pre-metric theories various subtleties can arise. In the uniaxial crystal studied in Sect.~\ref{sec:uniaxial}, for example, the Hamiltonian flow of~$\mathcal{G}(x,k)$ degenerates along the optic axis, necessitating a more ramified description of the wave-front set.
Evidently, the existence of states obeying the $\mu$SC places non-trivial restrictions on the wave-front set of $\sigma$.

Any two-point functions $\Lambda_2$, $\Lambda_2'$ satisfying the microlocal spectrum condition (even for distinct states) differ only in their smooth part.
To see this, set $u=\Lambda_2'-\Lambda_2$ and observe that, on one hand, $\WF(u)\subset\mathcal{N}^+\times\mathcal{N}^-$ by~\eqref{eq:microlocal}, while on the other, the generalized CCR ensures that $u$ is symmetric up to smooth errors.
Thus
\begin{equation*}
  \WF(u) = \WF(u^T) = \WF(u)\cap \WF(u^T)
  \subset\left(\mathcal{N}^+\times\mathcal{N}^-\right)\cap \left(\mathcal{N}^-\times\mathcal{N}^+\right) = \emptyset,
\end{equation*}
and we see that $u$ is smooth.

We also remark that a state~$\Lambda$ for $\mathfrak{A}$ induces by~\eqref{eq:smeared_F} a unique two-point function for the smeared field strengths, a second-rank covariant tensor bi-distribution which inherits the microlocal properties of $\Lambda_2$.
Moreover, the anti-symmetric part of this two-point function is fixed completely in terms of the restriction of $\sigma$ to vector densities and is therefore common to all states.

%******************************************************************************%
\section{Quantum energy inequality}
\label{sec:qei}

In this section we state and prove a QEI for pre-metric electrodynamics.
The proof follows the structure of~\cite{Fewster:1999gj,fewster:2003} but with some differences following from the more complicated form of both the energy density and the lightcone structure.
In addition, \cite{Fewster:1999gj,fewster:2003} established QEIs for averaging the energy density along timelike curves.
Here the structure which determines the `timelike' curves is the Fresnel polynomial and its dual polynomial.

We prove a QEI for curves $\gamma$ and their conormals $n$ satisfying two assumptions. First, the classical sWEC should hold along $\gamma$, \ie, the energy density is non-negative and vanishes precisely at points of vanishing field strength.
In Sect.~\ref{sub:qei:class_pt_split} this assumption will be used to construct a suitable point-split classical energy density in a `sum of squares' form.
Second, the trajectory must be `subluminal', which is equivalent to the fact that its tangent vector has everywhere positive contractions against every future-pointing null covector as proven in Sect.~\ref{sub:pmedyn:obs} (recall that there may be multiple lightcones that may touch or cross each other).
In Sect.~\ref{sub:qei:quant_pt_split}, this assumption will be used in the definition of the quantized point-split energy density.
Here, techniques from microlocal analysis are used.
Once this is done, the actual proof of the QEI in Sect.~\ref{sub:qei:proof} can follow established lines~\cite{Fewster:1999gj,fewster:2003}.

The setting has been kept as general as possible to accommodate variable constitutive laws -- even though the quantization in~\cite{Pfeifer:2016har} was worked out only in the constant case, the outline of the theory seems clear enough and what is lacking is a rigorous and general existence proof for the Pauli--Jordan propagator, so that a commutator may be defined.
In Sect.~\ref{sec:uniaxial} we will compute the QEI bound in detail for electrodynamics in a translationally invariant uniaxial birefringent crystal.
This will also show that averages of the energy density along `interluminal observer trajectories', whose tangents have positive contractions against some, but not all, null covectors, do not obey QEIs.

Finally, we remark that instead of providing an inequality for the energy density below, we could have produced one for either the electric or the magnetic field squared -- the general methods would have been the same.

%++++++++++++++++++++++++++++++++++++++++++++++++++++++++++++++++++++++++++++++%
\subsection{Classical point-split energy density}
\label{sub:qei:class_pt_split}

Let $\gamma:I\to M$ be a smooth curve, for some open interval $I\subset\RR$ and let $\rho$ be the energy density along~$\gamma$, defined as in Sect.~\ref{sub:pmedyn:edensity} with respect to a choice of frame~$e$ in a contractible neighbourhood~$\mathcal{T}$ of~$\gamma$.
If the sWEC holds on~$\gamma$ (with respect to~$e$) then the tensor fields~$\chi_1$ and~$\chi_2$ induce positive definite metrics along~$\gamma$ on the magnetic and electric subspaces respectively.
Hence this also holds within some neighbourhood of~$\gamma$, which we may take without loss of generality to be~$\mathcal{T}$ (redefining it if necessary).

We may therefore write
\begin{subequations}\label{eq:chi_12-6x6_repr}\begin{align}
  \chi_1^{abcd} F_{ab} F_{cd} &= X_1^{AB} \mathfrak{b}_A^{ab} \mathfrak{b}_B^{cd} F_{ab} F_{cd}, \\
  \chi_2^{abcd} F_{ab} F_{cd} &= X_2^{AB} \mathfrak{e}_A^{ab} \mathfrak{e}_B^{cd} F_{ab} F_{cd},
\end{align}\end{subequations}
where the indices $A,B$ run over $1,2,3$, $\mathfrak{b}_A^{ab}$ and $\mathfrak{e}_A^{ab}$ are smooth dual frames for the magnetic and electric subspaces respectively, and $X_1$ and $X_2$ are smooth families of real symmetric $3\times 3$ matrices.
Using the Kronecker-$\delta$ to raise and lower matrix indices, $(X_r)^{A}_{\phantom{A}B}$ ($r=1,2$) are positive definite (with respect to the inner product $\delta_{AB}$ on $3$-dimensional vectors) at each point of~$\mathcal{T}$ and have spectra uniformly bounded away from zero on compact subsets.
They therefore possess (unique) smooth\footnote{Square roots of uniformly positive definite matrices vary smoothly with the matrix as can be seen by an application of the inverse function theorem.
We are grateful to Simon Eveson for discussions on this matter.} positive square roots~$(Y_r)^{A}_{\phantom{A}B}$ on any such subset of~$\mathcal{T}$ and indeed, by considering a compact exhaustion, on all of~$\mathcal{T}$.
Accordingly, $X_r^{AB}=(Y_r)^{A}_{\phantom{A}C}(Y_r)^{C}_{\phantom{C}D}\delta^{DB}$ and we have a sum-of-squares form for the electromagnetic energy density
\begin{equation}\label{eq:rhoclass}
  \rho = \frac12 \delta_{AB} (\mathfrak{E}^A \mathfrak{E}^B + \mathfrak{B}^A   \mathfrak{B}^B),
\end{equation}
where
\begin{equation*}
  \mathfrak{E}^B = \frac12 F_{ab} \mathfrak{e}_A^{ab} Y_1^{AB},
  \quad
  \mathfrak{B}^B = \frac12 F_{ab} \mathfrak{b}_A^{ab} Y_2^{AB}
\end{equation*}
are linear combinations of the components of $F_{ab}$ with smoothly varying real coefficients.
The notation $\mathfrak{E}^B$ and $\mathfrak{B}^B$ is intended to remind the reader of `electric' and `magnetic'.
However, we caution that in general these quantities are not to be interpreted as components of the electric or magnetic field strengths.
Indeed, the energy density in pre-metric electrodynamics involves the electric and magnetic field strengths and also the electric and magnetic excitations (see, \eg,~\cite[\S B.5.3]{Hehl}).
Thus, our quantities~$\mathfrak{E}^B$ and~$\mathfrak{B}^B$ are combinations of field strengths and excitations.

The expression~\eqref{eq:rhoclass} allows us to define the classical \emph{point-split energy density} for a worldline~$\gamma(\tau)$ as
\begin{equation}\label{eq:classPS}
  \rho(\tau, \tau')
  \defn \frac12 \delta_{AB} \bigl( \mathfrak{E}^A(\gamma(\tau)) \mathfrak{E}^B(\gamma(\tau')) + \mathfrak{B}^A(\gamma(\tau)) \mathfrak{B}^B(\gamma(\tau')) \bigr).
\end{equation}
It is obvious that $\rho(\tau,\tau)=\rho(\tau)$ is again the energy density with respect to the worldline~$\gamma$ and the frame $e$.

There exists an interesting class of constitutive laws and worldlines for which the point-split energy density can be obtained without the need to use the square-root construction above.
Namely, suppose that there exists a global Cartesian coordinate system on spacetime in which the components of the constitutive density is translationally invariant, and consider an inertial worldline $\gamma$ with an associated framing $e$ that is translationally invariant.
Then the density factor~$\varepsilon(e)^{-1}$ is constant and the point-split energy density may be given in the form
\begin{align}\label{eq:psplitconst}
	\rho(\tau, \tau')
	&= \frac18 (\chi^{abcd}_1 + \chi^{abcd}_2) F_{ab}(\gamma(\tau)) F_{cd}(\gamma(\tau'))\nonumber\\
	&= \frac18 \varepsilon(e)^{-1} (\chi^{abcd} - 2 \chi^{ebcd} n_e \dot\gamma^a - 2 \chi^{abed} n_e \dot\gamma^c) F_{ab}(\gamma(\tau)) F_{cd}(\gamma(\tau')).
\end{align}
Observe that translational invariance of $\chi$ and $e$ are necessary to obtain this expression because otherwise one would have to specify where the prefactors before the field strengths were evaluated.
The system of inertial worldlines in an uniaxial crystal, which we will discuss in detail in Sect.~\ref{sec:uniaxial}, belongs to this class.

%++++++++++++++++++++++++++++++++++++++++++++++++++++++++++++++++++++++++++++++%
\subsection{Quantized point-split energy density}
\label{sub:qei:quant_pt_split}

The classical fields $\mathfrak{E}^A$ and $\mathfrak{B}^A$ are easily quantized.
For any (scalar) density $f$ compactly supported in $\mathcal{T}$, we define (for $B=1,2,3$)
\begin{equation*}
  \what{\mathfrak{E}}^B(f) = \frac12 \what{F}(\mathfrak{e}_A Y_1^{AB} f), \quad
  \what{\mathfrak{B}}^B(f) = \frac12 \what{F}(\mathfrak{b}_A Y_2^{AB} f)
\end{equation*}
which are elements of the algebra $\mathfrak{A}$.
Any (sufficiently regular) state~$\Lambda$ induces scalar bi\hyp{}distributions $\mathfrak{E}_{2,\Lambda}^{AB},\mathfrak{B}_{2,\Lambda}^{AB} \in \mathscr{D}'(\mathcal{T}\times \mathcal{T})$ by
\begin{equation*}
  \mathfrak{E}_{2,\Lambda}^{AB}(f_1,f_2) \defn \Lambda\bigl(\what{\mathfrak{E}}^A(f_1)\what{\mathfrak{E}}^B(f_2)\bigr),
  \quad
  \mathfrak{B}_{2,\Lambda}^{AB}(f_1,f_2) \defn \Lambda\bigl(\what{\mathfrak{B}}^A(f_1)\what{\mathfrak{B}}^B(f_2)\bigr),
\end{equation*}
whose wave-front sets are both contained in $\mathcal{N}^+\times\mathcal{N}^-$ if $\Lambda$ obeys the microlocal spectrum condition.

We note two important properties of the distributions $\mathfrak{E}_{2,\Lambda}^{AA}$ and $\mathfrak{B}_{2,\Lambda}^{AA}$ (no sum on $A$): (a) their antisymmetric parts are independent of the state $\Lambda$, being determined by the CCRs; (b) they are of positive type as a consequence of the positivity of $\Lambda$ as a state.

Our aim is to define the (un-renormalized) point-split energy density along $\gamma$ in state $\Lambda$ as a pull-back
\begin{equation*}
  \rho_\Lambda = \frac{1}{2}\delta_{AB}\varphi^*  \bigl(\mathfrak{E}_{2,\Lambda}^{AB} + \mathfrak{B}_{2,\Lambda}^{AB}\bigr),
\end{equation*}
where
\begin{align*}
  \varphi: I\times I &\rightarrow M\times M \nonumber\\
  (\tau, \tau') &\mapsto \varphi(\tau,\tau')=(\gamma(\tau),\gamma(\tau')).
\end{align*}
The required pull-back exists provided that $\gamma$ is a subluminal trajectory as we now describe.
Note that in this case we have $k \cdot \dot\gamma > 0$ for all $k\in\mathcal{N}^+$.
This condition implies directly that the intersections $N_\gamma \cap \mathcal{N}^\pm$ are empty, where
\begin{equation*}
  N_\gamma = \bigl\{(\gamma(\tau),k)\in T^*\!M \;\big|\; k \cdot \dot\gamma(\tau) = 0 \bigr\}
\end{equation*}
is the set of conormals of $\gamma$, and also that the pull-backs
\begin{equation*}
  \gamma^*\mathcal{N}^\pm = \bigl\{(\tau,k \cdot \dot\gamma(\tau))\in T^*I \;\big|\; (\gamma(\tau),k)\in\mathcal{N}^\pm  \bigr\}
\end{equation*}
are contained in $I\times\RR^\pm \subset T^*I$.
Now the conormals of the map~$\varphi$ are the same as stated in~\cite{Fewster:1999gj}
\begin{align*}
	N_\varphi = \bigl\{ (\gamma(\tau),k; \gamma(\tau'), k')\in T^*(M\times M) \;\big|\; k \cdot \dot\gamma(\tau) = k' \cdot \dot\gamma(\tau') = 0 \bigr\} = N_\gamma\times N_\gamma
\end{align*}
and we deduce immediately that $(\mathcal{N}^+\times\mathcal{N}^-)\cap N_\varphi$ is empty.
By the microlocal spectrum condition and H\"ormander's criterion~\cite[Thm 2.5.11${}'$]{Hoermander_FIOi:1971}, it follows that the pull-backs $\varphi^* \mathfrak{E}_{2,\Lambda}^{AB}$ and $\varphi^* \mathfrak{B}_{2,\Lambda}^{AB}$ exist as distributions in $\mathscr{D}'(\RR\times\RR)$, with wave-front sets obeying
\begin{equation*}
  \WF(\varphi^* \mathfrak{E}_{2,\Lambda}^{AB}),~\WF(\varphi^* \mathfrak{B}_{2,\Lambda}^{AB})
  \subset \varphi^*(\mathcal{N}^+\times\mathcal{N}^-)\subset (I\times I)\times (\RR^+\times\RR^-)
\end{equation*}
in $T^*(I\times I)$.
Furthermore, the distributions $\varphi^* \mathfrak{E}_{2,\Lambda}^{AA}$ and $\varphi^* \mathfrak{B}_{2,\Lambda}^{AA}$ (no sum) inherit the properties of having state-independent antisymmetric parts and being of positive type.
Consequently, the point-split energy density $\rho_\Lambda$ exists, is of positive type, and has wave-front set
\begin{equation}\label{eq:WFsetbounds}
  \WF(\rho_\Lambda) \subset  (I\times I)\times (\RR^+\times\RR^-)
\end{equation}
in $T^*(I\times I)$.

\begin{figure}
	\centering
  \begin{tikzpicture}[scale=0.6]
    \draw [dashed,fill=gray!15] (2.92,2.86) -- (0,0) -- (-2.92,2.86) -- cycle;
    \begin{scope}
      \clip (0,3) ellipse (5 and 1);
      \draw [thick,fill=gray!30] (2.92,2.86) -- (0,0) -- (-2.92,2.86) -- cycle;
    \end{scope}
    \draw [thick,fill=gray!30] (0,3) node[xshift=-5mm] {$\Gamma$} ellipse (3 and .6);

    \draw [thick] (4.63,2.62) -- (0,0) -- (-4.63,2.62);
    \draw [thick] (0,3) ellipse (5 and 1);

    \fill (0,0) circle (0.1) node[below] {$x$};

    \node at (-3.5,0) {$T^*_x M$};
  	\node[scale=.8,rotate=8] at (-3.2,4.1) {$\zeta^{-1}(k,k)=0$};
  	\node[scale=.8,rotate=5] at (1.4,2.85) {$\eta^{-1}(k,k)=0$};
  \end{tikzpicture}
  \quad
  \begin{tikzpicture}[scale=0.6]
    \draw [dashed] (2.92,2.86) -- (0,0) -- (-2.92,2.86);
    \draw [color=red,thick,dashed,->] (0,0) -- (3.9,3.15);
    \begin{scope}
      \clip (0,3) ellipse (5 and 1);
      \draw [thick] (2.92,2.86) -- (0,0) -- (-2.92,2.86);
      \draw [color=red,very thick,->] (0,0) -- (3.9,3.15);
    \end{scope}

    \draw [color=blue,thick,dashed,->] (0,0) -- (-1.2,3.15);
    \begin{scope}
      \clip (0,3) ellipse (3 and .6);
      \draw [color=blue,very thick,->] (0,0) -- (-1.2,3.15);
    \end{scope}
    \draw [thick] (0,3) ellipse (3 and .6);

    \draw [thick] (4.63,2.62) -- (0,0) -- (-4.63,2.62);
    \draw [thick] (0,3) ellipse (5 and 1);

    \fill (0,0) circle (0.1) node[below] {$x$};

    \node at (-3.5,0) {$T_x M$};
  	\node[scale=.8,rotate=8] at (-3.2,4.1) {$\eta(X,X)=0$};
  	\node[scale=.8,rotate=5] at (1.4,2.8) {$\zeta(X,X)=0$};
  \end{tikzpicture}
  \caption{Cone structure in the cotangent (left) and tangent (right) space at $x$ on spacetime. Subluminal directions are inside the inner cone, interluminal directions lie between the cones, and superluminal directions outside of both cones in $T_xM$.}
	\label{fig:bimetric}
\end{figure}
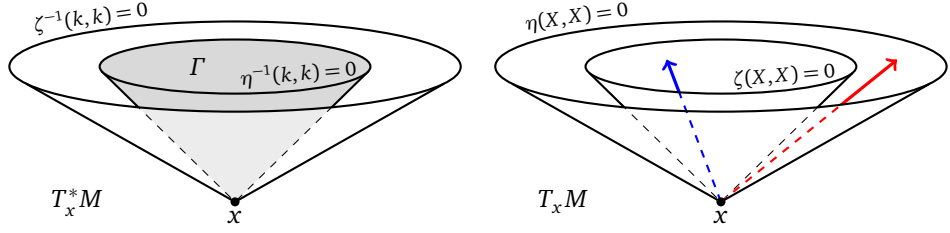

It is useful to illustrate the above in the example of a \emph{bi-metric Fresnel polynomial}, for which $\mathcal{G}(k) = \vartheta\, \zeta^{-1}(k,k)\, \eta^{-1}(k,k)$, where $\zeta^{-1}$ and $\eta^{-1}$ are the inverses of two Lorentzian metrics with signature  $\mathord{-}\mathord{+}\mathord{+}\mathord{+}$ and $\vartheta$ is a density, and we assume that the (positive frequency) lightcone of $\eta^{-1}$ lies inside that of $\zeta^{-1}$ in the cotangent space $T_x^{\mathrlap{*}}M$ (see Fig.~\ref{fig:bimetric}).
Of course this means that the (future) lightcone of $\zeta$ lies within that of $\eta$ in the tangent space $T_xM$.
(Uniaxial birefringent crystals have constitutive laws of this type, in a degenerate case where the two lightcones touch along two generators.)
In this situation, $\mathcal{N}^+$ consists of the union of the positive frequency lightcones of both $\eta^{-1}$ and $\zeta^{-1}$ in $T_x^{\mathrlap{*}}M\setminus\{0\}$.
Meanwhile, the hyperbolicity double cone of the bi-metric theory consists of the covectors satisfying $\eta^{-1}(k,k)<0$, from which one component can be chosen to be the hyperbolicity cone  $\Gamma\subset T_x^{\mathrlap{*}}M$ defining the causal orientation.

Vectors which are timelike with respect to both metrics, like the blue vector in Fig.~\ref{fig:bimetric}, have positive contraction with all positive frequency null covectors and are tangents to subluminal trajectories.
They form the cone $\Gamma^\#$, which consists of the future timelike vectors of~$\zeta$ in $T_xM$.
As described above, the point-split energy density can be defined for these curves.
On the other hand, interluminal vectors are timelike with respect to one of the metrics while being spacelike for the other, like the red vector in Fig.~\ref{fig:bimetric}.
They have positive contractions with some positive frequency null covectors but not with others, and so the above construction does not apply.
This is not to say that the point-split energy density does not exist as a distribution along interluminal trajectories, but rather that it fails the sufficient condition provided by microlocal techniques.
Thus the bounds~\eqref{eq:WFsetbounds} (and also the QEI below) are not guaranteed to hold.
Indeed, we will show later that, for uniaxial birefringent crystals, there are states of the QFT that violate the QEIs along interluminal trajectories.

%++++++++++++++++++++++++++++++++++++++++++++++++++++++++++++++++++++++++++++++%
\subsection{Statement and proof of the QEI}
\label{sub:qei:proof}

The discussion above involved two assumptions on the curve $\gamma:I\to M$ and the constitutive law.
The first was that the classical sWEC holds along $\gamma$, with respect to a choice of frame $e$, allowing the construction of a point-split energy density on $M \times M$, while the second required $\gamma$ to be a subluminal trajectory, thus allowing the point-split energy density to be defined using a pull-back.

With these assumptions in force, let $\Omega$ be a fixed reference state on $\mathfrak{A}$ obeying the microlocal spectrum condition.
In particular, if $\Lambda$ is also a state obeying the microlocal spectrum condition, the differences $\mathfrak{E}_{2,\Lambda}^{AB}- \mathfrak{E}_{2,\Omega}^{AB}$ and $\mathfrak{B}_{2,\Lambda}^{AB}- \mathfrak{B}_{2,\Omega}^{AB}$ are smooth, and therefore the same is true of $\rho_\Lambda-\rho_\Omega$.
Furthermore, $\rho_\Lambda-\rho_\Omega$ is symmetric in its arguments, because the antisymmetric part of $\rho_\Lambda$ is state-independent.
The expectation value in $\Lambda$ of the normal ordered energy density on $\gamma$ (relative to the reference state $\Omega$) may then be defined as
\begin{equation*}
  \langle \norder{\rho(\tau)} \rangle_\Lambda \defn (\rho_\Lambda-\rho_\Omega)(\tau,\tau).
\end{equation*}
With our preparations complete, the QEI may be stated.
It asserts that the inequality
\begin{align}\label{eq:QEI}
	\int_\RR g(\tau)^2 \langle \norder{\rho(\tau)} \rangle_\Lambda\, \dif\tau
	&\geq -\frac1\uppi \int_0^\infty \left( \iint_{\RR^2} g(\tau) g(\tau') \e^{-\im \beta(\tau-\tau')} \rho_\Omega(\tau,\tau')\, \dif\tau\, \dif\tau' \right) \dif\beta
\end{align}
holds for all states $\Lambda$ on $\mathfrak{A}$ obeying the microlocal spectrum condition, and all real-valued compactly supported $g \in C_0^\infty(I)$, and that the right-hand side of the inequality is finite.
Note that this lower bound is independent of the state $\Lambda$ but depends on the reference state $\Omega$.

The proof is similar to those of~\cite{Fewster:1999gj,fewster:2003} -- the main differences are contained in the construction of the energy density and its pull-back.
Therefore, the following argument will be kept brief.
The first step is to split the points apart, by insertion of a $\delta$-function in its Fourier representation
\begin{align*}\MoveEqLeft
	\int_\RR g(\tau)^2 \langle \norder{\rho(\tau)} \rangle_\Lambda\, \dif\tau \\
	&= \frac{1}{2\uppi} \int_\RR \left( \iint_{\RR^2} g(\tau) g(\tau') \e^{-\im \beta (\tau-\tau')} \bigl(\rho_\Lambda(\tau,\tau') - \rho_\Omega(\tau,\tau')\bigr)\, \dif\tau\, \dif\tau' \right) \dif\beta.
\end{align*}
Next, we exploit the symmetry of $\rho_\Lambda-\rho_\Omega$ to write the expression as an integral over~$\RR^+$:
\begin{align*}
  \text{L.H.S.}	&= \frac1\uppi \int_0^\infty \left( \iint_{\RR^2} g(\tau) g(\tau') \e^{-\im \beta (\tau-\tau')} \bigl( \rho_\Lambda(\tau,\tau') - \rho_\Omega(\tau,\tau') \bigr)\, \dif\tau\, \dif\tau' \right) \dif\beta \nonumber\\
  &\geq -\frac1\uppi \int_0^\infty \left( \iint_{\RR^2} g(\tau) g(\tau') \e^{-\im \beta(\tau-\tau')} \rho_\Omega(\tau,\tau')\, \dif\tau\, \dif\tau' \right) \dif\beta \nonumber\\
  &\geq -\frac1\uppi \int_0^\infty \mathcal{F}\bigl( (g\otimes g)\rho_\Omega \bigr)(-\beta,\beta)\, \dif\beta,
\end{align*}
where we have also used the positive type property to discard $\rho_\Lambda$ and written the resulting expression as an integral of a Fourier transform, denoted here by $\mathcal{F}$.
Finally, \eqref{eq:WFsetbounds} entails that the Fourier transform of the compactly supported distribution $(g\otimes g)\rho_\Omega$ decays faster than any inverse power along $(-\beta,\beta)$ as $\beta\to\infty$.
Thus the integral is finite and the QEI is established.

%******************************************************************************%
\section{Uniaxial birefringent crystals}
\label{sec:uniaxial}

In order to gain a deeper understanding of how QEIs in pre-metric electrodynamics differ from those in ordinary Maxwell theory, we consider a simple translationally invariant birefringent crystal in Minkowski spacetime whose constitutive density takes the following form in a global Cartesian coordinate system \cite{Perlick,Pfeifer:2016har}
\begin{equation}\label{eq:constitutive}
  \chi^{abcd} = \abs{\eta}^\frac12 (2 \eta^{c[a} \eta^{b]d} + 4 X^{[a} U^{b]} X^{[d} U^{c]}).
\end{equation}
Here we use the following notation:
\begin{itemize}
  \item $\eta$ is the Minkowski metric (with signature $\mathord{-}\mathord{+}\mathord{+}\mathord{+}$) and $\abs{\eta}^\frac12$ is the associated density.
  When we raise or lower indices, we use this metric.
  \item $U$ is a timelike vector field (normalized such that $\eta(U,U)=-1$), which represents the rest frame of the crystal.
  \item $X$ is a spacelike vector orthogonal to $U$ (\viz, $\eta(X,U) = 0$), which defines the optic axis of the birefringent crystal.
  We set $\xi^2 = \eta(X,X)$.
\end{itemize}
The quantities $\eta$, $X$ and $U$ and hence $\chi$ are all translationally invariant; in particular, $\chi^{abcd}$ is constant in the chosen global Cartesian coordinate system and $\abs{\eta}=1$.
We label the coordinate with indices running from $0$ to $3$ such that $0$ indicates a timelike direction.
We can always rotate the coordinate system such that $U = U^a \partial_a = \partial_0$ and $X = X^a \partial_a = \xi \partial_1$.
From now on we will use this specific coordinate system in our calculations.

%++++++++++++++++++++++++++++++++++++++++++++++++++++++++++++++++++++++++++++++%
\subsection{Lightrays}

The Fresnel polynomial of the constitutive law~\eqref{eq:constitutive} is bi-metric,
\begin{equation}\label{eq:Fresnel-uni}
  \mathcal{G}(k) = \abs{\eta}^\frac12 \eta^{-1}(k,k) \zeta^{-1}(k,k)
\end{equation}
where $\eta^{ab}=(\eta^{-1})^{ab}$ and
\begin{equation*}
  (\zeta^{-1})^{ab} = \eta^{ab} - \xi^2 U^a U^b + X^a X^b
\end{equation*}
are components of inverse metrics while $\eta_{ab}$ and
\begin{align}\label{eq:zeta}
  \zeta_{ab} = \eta_{ab} + \frac{\xi^2}{1+\xi^2}U_a U_b - \frac{1}{1+\xi^2} X_a X_b
\end{align}
denote the components of the metrics.

The zeros of the Fresnel polynomial determine the propagation of light rays in the geometrical optics approximation.
In the case at hand it they are given by the \emph{ordinary lightcone} defined through
\begin{equation*}
  \eta^{-1}(k,k) = 0,
\end{equation*}
and the \emph{extraordinary lightcone} described by
\begin{equation*}
  \zeta^{-1}(k,k) = 0.
\end{equation*}
Examples of extraordinary and ordinary lightrays are $k_a \propto (1,0,0,(1+\xi^2)^{1/2})$ and $k_a \propto (1,0,0,1)$, respectively.
There is one direction along which the ordinary and extraordinary lightcones coincide; namely the optic axis $X$.
If both $\eta^{-1}(k,k) = 0$ and $\zeta^{-1}(k,k) = 0$, then $(k \cdot X)^2 = \xi^2 (k \cdot U)^2$, which occurs precisely when
\begin{equation*}
  k_a \propto X_a \pm \xi U_a,
\end{equation*}
\ie, in coordinates $k_a \propto (\pm 1,1,0,0)$.
The ordinary lightcone given by $\eta$ is the inner lightcone in the cotangent spaces, while, by duality, it is the outer lightcone in the tangent spaces, compare Fig.~\ref{fig:bimetric}.
This inner cone is also the hyperbolicity cone according to the discussion in Sect.~\ref{sub:pmedyn:fresnel}.
Thus the ordinary lightcone determines maximum velocity of light in the medium.
In our terminology, an observer whose velocity lies within the extraordinary lightcone is subluminal, and one whose velocity lies between the extraordinary and ordinary lightcones is interluminal.

In a $1+3$ split, we can write each momentum covector $k$ as $k = (k_0, \vec{k})$.
For each fixed $\vec{k}$ we define $\omega(\vec{k})$ and $\tilde\omega(\vec{k})$ as the unique positive zeros of $\eta(k,k)^{-1}$ and $\zeta(k,k)^{-1}$, as functions of $k_0$, namely
\begin{subequations}\label{eq:freq}\begin{align}
  \omega(\vec{k}) &= \sqrt{k_1^2+k_2^2+k_3^2},\\
  \tilde\omega(\vec{k}) &= \sqrt{k_1^2 + (k_2^2+k_3^2)/(1+\xi^2)}.
\end{align}\end{subequations}
Clearly, $\omega = \omega(\vec{k})$ and $\tilde\omega = \tilde\omega(\vec{k})$ are the frequencies of a lightray with momentum~$\vec{k}$, propagating on the ordinary or extraordinary (forward) lightcone, measured by an observer at rest with respect to the crystal.
In the following we will suppress the explicit $\vec{k}$-dependence of~$\omega$ and~$\tilde\omega$.

%++++++++++++++++++++++++++++++++++++++++++++++++++++++++++++++++++++++++++++++%
\subsection{Green functions, two-point functions and the Pauli--Jordan propagator}
\label{sub:uniaxial:propagators}

In this subsection, we compute the Pauli--Jordan propagator, which is necessary to construct the algebra $\mathfrak{A}$ of smeared quantum fields, and present a suitable reference state on $\mathfrak{A}$ that satisfies the microlocal spectrum condition. Both the Pauli--Jordan propagator and the two-point function of the state may be obtained from an analysis of the retarded and advanced Green functions $E^{\text{ret/adv}}_{ab}(x, x')$ using contour integral methods.

As stated in Sect.~\ref{sub:pmedyn:fresnel}, the quasi-inverse $\mathcal{E}$ of the principal symbol $\mathcal{M}$ is the main ingredient in the construction of the Green functions.
The results of~\cite{Pfeifer:2016har} give
\begin{equation*}
  E^{\text{ret/adv}}_{ab}(x, x') \defn \lim_{\varepsilon \to 0^+} \frac{1}{(2\uppi)^4} \int_{\RR^3} \int_{\RR\pm\im\varepsilon} \mathcal{E}_{ab}(k)\, \e^{-\im \vec{k} \cdot (\vec{x} - \vec{x}') - \im k_0 (t - t')}\, \dif k_0\, \dif \vec{k},
\end{equation*}
where the limit in $\varepsilon$ is taken in the sense of distributions and, as before,
\begin{equation*}
	\mathcal{E}_{ab}(k) = \frac{\mathcal{Q}_{cd}(k) \pi^c_a(k) \pi^b_d(k)}{\mathcal{G}(k)}.
\end{equation*}
Since $\mathcal{Q}$ is contracted with the projectors~$\pi$, it can be replaced with a tensor~$\tilde{\mathcal{Q}}$ which differs from~$\mathcal{Q}$ by terms proportional to the momentum covector~$k$.
For the uniaxial crystal, a suitable~$\tilde{\mathcal{Q}}$ is given by, see Appx.~A in~\cite{Pfeifer:2016har},
\begin{align}\label{eq:tildeQ}
	\tilde{\mathcal{Q}}_{ab}(k)
	&\defn \eta_{ab} \zeta^{-1}(k,k) + q_a(k)q_b(k)
\end{align}
with
\begin{equation*}
  q_a(k) \defn (k \cdot X) U_a - (k \cdot U) X_a.
\end{equation*}
Eq.~\eqref{eq:tildeQ} nicely demonstrates the effect of the crystal compared to Maxwell vacuum electrodynamics.
Indeed, the properties of the crystal are encoded in the vector $q$, while the first term is the same as in vacuum electrodynamics.

Defining the operators $E^{\text{ret/adv}}$ by
\begin{equation*}
  (E^{\text{ret/adv}}j)_a(x) \defn \int_{\mathbb{R}^4} \, E^{\text{ret/adv}}_{ab}(x, x') j^b(x')\, \dif x',
\end{equation*}
it may be verified that the support of $E^{\text{ret/adv}}j$ is contained in the causal future/past of the support of $j$.
In this context the notions of causal future/past refer to the causal structure defined by the Fresnel polynomial, see \cite[Sect.~2.3]{Pfeifer:2016har} for the technical definitions. Briefly, the causal future of a point $x\in M$ comprises all points which can be connected to $x$ by future/past-pointing non-superluminal curves, \ie, all curves with future/past-pointing subluminal, interluminal or luminal tangent as they are defined in Sect.~\ref{sub:pmedyn:obs}.

The important property of the Green functions is that they are inverses to the differential operator $P$ up to gauge: that is,
the identities
\begin{align*}
P E^{\text{ret/adv}}j &= j \\
E^{\text{ret/adv}} PA &= A+\dif\lambda
\end{align*}
hold for all conserved compactly supported vector densities $j$, and for any compactly supported $1$-form $A$, where $\lambda$ is a smooth function depending on $A$ and the choice of advanced or retarded.

The Pauli--Jordan propagator is the difference of the advanced and retarded Green functions:
\begin{equation*}
  \upDelta_{ab}(x, x') \defn E_{ab}^{\text{adv}}(x, x') - E_{ab}^{\text{ret}}(x, x').
\end{equation*}
We will use residue methods to evaluate the $k_0$-integrals involved in defining $\upDelta_{ab}(x, x')$, and therefore the gauge fixing vector field $\kappa^a(k)$ (used to construct the projectors $\pi$ in $\mathcal{E}_{ab}$) must be defined on $k\in \CC\times\RR^3$, depend meromorphically on $k_0$ and obey $k\cdot\kappa(k)=1$ everywhere except at its poles. Noting that $q^a k_a=0$, a convenient choice is given by
\begin{equation*}
  \kappa^a = \frac{k^a+\im q^a}{\eta^{-1}(k,k)},
\end{equation*}
for which one may calculate, using $\zeta^{-1}(k,k) = \eta^{-1}(k,k)-\eta^{-1}(q,q)$, that
\begin{equation*}
  \mathcal{E}_{ab}(k) = \frac{\eta_{ab}}{\eta^{-1}(k,k)} - \frac{(k_a+\im q_a)(k_b+\im q_b)}{\zeta^{-1}(k,k)\eta^{-1}(k,k)}.
\end{equation*}
Evidently, the poles of $\mathcal{E}_{ab}(k)$ in $k_0$ for fixed $\vec{k}$ are precisely at $k_0=\pm\omega$ and $k_0=\pm\tilde\omega$.
Note that $\kappa$ is complex even when $k$ is restricted to the real axis; consequently, $\Delta_{ab}(x, x')$ also becomes complex.
The imaginary part is associated with pure gauge terms and therefore has no physical significance.

\begin{figure}
  \centering
  \begin{tikzpicture}
    % axes
    \draw[->] (-3,0) -- (3,0);
    \draw[->] (0,-1.7) -- (0,1.7);

    % poles
    \node[circle,fill,inner sep=1pt,label=above:$\mkern-8mu+\tilde\omega$] at (+.8,0){};
    \node[circle,fill,inner sep=1pt,label=above:$\mkern-8mu-\tilde\omega$] at (-.8,0){};
    \node[circle,fill,inner sep=1pt,label=above:$\mkern-8mu+\omega$] at (+1.7,0){};
    \node[circle,fill,inner sep=1pt,label=above:$\mkern-8mu-\omega$] at (-1.7,0){};

    % contour C with arrows
    \draw[very thick,decoration={markings,mark=at position 0.14 with {\arrow{>}},mark=at position 0.37 with {\arrow{>}},mark=at position 0.14 with \node[above]{$C$};,mark=at position 0.64 with {\arrow{>}},mark=at position 0.87 with {\arrow{>}}},postaction={decorate}] (0,0) ellipse (2.5 and 1.3);

    % contour C^+ with arrows
    \draw[very thick,decoration={markings,mark=at position 0.37 with {\arrow{>}},mark=at position 0.37 with \node[above]{$C^+$};,mark=at position 0.87 with {\arrow{>}}},postaction={decorate}] (1.25,0) ellipse (1.0 and 0.75);

    % contour C^- with arrows
    \draw[very thick,decoration={markings,mark=at position 0.14 with {\arrow{>}},mark=at position 0.14 with \node[above]{$C^-$};,mark=at position 0.64 with {\arrow{>}}},postaction={decorate}] (-1.25,0) ellipse (1.0 and 0.75);
  \end{tikzpicture}
  \caption{Illustration of the contours $C$, $C^+$ and $C^-$ used to compute the Pauli--Jordan propagator and the positive and negative frequency bi-distributions.}
  \label{fig:contour-PJ}
\end{figure}
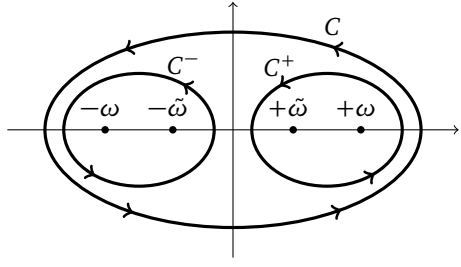

With these choices, $\Delta_{ab}(x,x')$ may be expressed as a contour integral
\begin{align}\label{eq:contour-PJ}
  \upDelta_{ab}(x, x')
  &= \frac{1}{(2\uppi)^4} \int_{\RR^3} \int_{C} \mathcal{E}_{ab}(k)\, \e^{-\im \vec{k} \cdot (\vec{x} - \vec{x}') - \im k_0 (t - t')}\, \dif k_0\, \dif \vec{k},
\end{align}
where the contour $C$ can depend on $\vec{k}$, provided it encircles all the poles once in the counterclockwise direction as shown in Fig.~\ref{fig:contour-PJ}.
Convergence of this integral is understood in the distributional sense; that is, the $k$-integrals should be taken after integrating against compactly supported vector densities in the two spacetime variables $x$ and $x'$.
This controls the integration in the large momentum limit and leaves only the question of possible divergences at finite momenta.

By changing the contour $C$ in \eqref{eq:contour-PJ} we may obtain distributions that will determine the positive and negative frequency two-point functions $\Delta^\pm$ of a vacuum state on the algebra of observables $\mathfrak{A}$.
Specifically, the kernel $\mp\im\Delta^\pm_{ab}(x, x')$ is obtained by using the contours $C^\pm$, which encircle only the positive ($+$) or negative ($-$) frequency poles (see Fig.~\ref{fig:contour-PJ}), instead of $C$.
It is immediately clear that
\begin{align}\label{eq:gCCR}
  \im\upDelta_{ab}(x,x') &= \upDelta^+_{ab}(x,x') - \upDelta^-_{ab}(x,x').
\end{align}

Avoiding for the moment the special case $k_2=k_3=0$ (in which case there are second order poles at $k_0=\pm\omega=\pm\tilde\omega$) the poles in the integrand are all first order and one easily computes, using $\eta^{-1}(k,k)=\omega^2-k_0^2$ and $\zeta^{-1}(k,k)=(1+\xi^2)(\tilde\omega^2-k_0^2)$, that
\begin{equation*}
  \operatorname*{res}_{k_0 =\omega} \mathcal{E}_{ab}(k_0, \vec{k}) = -\frac{\mathcal{U}_{ab}(\vec{k})}{2\omega}, \qquad
  \operatorname*{res}_{k_0 =\tilde\omega} \mathcal{E}_{ab}(k_0, \vec{k}) = -\frac{\tilde{\mathcal{U}}_{ab}(\vec{k})}{2\tilde\omega},
\end{equation*}
where
\begin{align}
  \mathcal{U}_{ab}(\vec{k}) &\defn \eta_{ab} + \frac{(k_a+i q_a)(k_b + i q_b)}{\eta^{-1}(q,q)}\bigg|_{k_0 = \omega}   \label{eq:Uab1}
  \\
  \tilde{\mathcal{U}}_{ab}(\vec{k}) &\defn - \frac{ (k_a+i q_a)(k_b + i q_b)}{(1+\xi^2)\eta^{-1}(q,q)} \bigg|_{k_0 = \tilde\omega} . \label{eq:Uab2}
\end{align}
Using the symmetry $\mathcal{E}_{ab}(-k) = \mathcal{E}_{ab}(k)$, one also has
\begin{equation*}
  \operatorname*{res}_{k_0 =-\omega} \mathcal{E}_{ab}(k_0, -\vec{k}) = \frac{\mathcal{U}_{ab}(\vec{k})}{2\omega}, \qquad
  \operatorname*{res}_{k_0 =-\tilde\omega} \mathcal{E}_{ab}(k_0, -\vec{k}) = \frac{\tilde{\mathcal{U}}_{ab}(\vec{k})}{2\tilde\omega}.
\end{equation*}
In the above calculations, we have used the equalities
\begin{align*}
  \zeta^{-1}(k,k) &= -\eta^{-1}(q(k),q(k))
\intertext{for $k$ on the ordinary lightcone $k_0 = \pm\omega$ and}
  \eta^{-1}(k,k) &= \eta^{-1}(q(k),q(k))
\end{align*}
for $k$ on the extraordinary lightcone $k_0 = \pm\tilde\omega$.

Assembling these results (and changing variables $\vec{k}\mapsto -\vec{k}$ for $\Delta^-_{ab}$) we find
\begin{align}\label{eq:posneg_freq}
  \upDelta^\pm_{ab}(x, x') \defn \frac{1}{(2\uppi)^3} \int_{\RR^3} \biggl( \mathcal{U}_{ab}(\vec{k}) \frac{\e^{\mp\im \omega (t-t')}}{2 \omega} + \tilde{\mathcal{U}}_{ab}(\vec{k}) \frac{\e^{\mp\im \tilde\omega (t-t')}}{2 \tilde\omega} \biggr)\, \e^{\mp\im \vec{k} \cdot (\vec{x} - \vec{x}')}\, \dif\vec{k}.
\end{align}
Note that the individual terms in the integrand have divergences as $k_2^2+k_3^2\to 0$, which is the special case in which the ordinary and extraordinary lightcone touch.
However, their sum remains regular in this limit, and the limiting value can be obtained by taking a residue at the double pole formed when the two single poles merge.
Thus the above integrals are well-defined in the distributional sense.
The Pauli--Jordan propagator is then
\begin{align}\label{eq:spatial-PJ}
  \upDelta_{ab}(x, x')
  &= \begin{aligned}[t]-\frac{1}{(2\uppi)^3} \int_{\RR^3} \Biggl(& \mathcal{U}_{ab}(\vec{k}) \frac{\sin \left(\omega (t-t')+\vec{k}\cdot(\vec{x}-\vec{x}')\right)}{\omega} \\
  &\mathllap{+\mskip\medmuskip} \tilde{\mathcal{U}}_{ab}(\vec{k}) \frac{\sin \left(\tilde\omega (t-t')+\vec{k}\cdot(\vec{x}-\vec{x}')\right)}{\tilde\omega}\Biggr)\, \dif\vec{k}.\end{aligned}
\end{align}
Since $\mathcal{U}_{ab}$ and $\tilde{\mathcal{U}}_{ab}$ are symmetric in their indices, we have
\begin{equation*}
  \upDelta_{ab}(x,x')=\upDelta_{(ab)}(x,x')
  \quad\text{and}\quad
  \upDelta^\pm_{ab}(x,x')=\upDelta^\pm_{(ab)}(x,x').
\end{equation*}
It is also clear that $\upDelta^-_{ab}(x,x')=\upDelta^+_{ba}(x',x)$, and hence $\upDelta_{ab}(x',x)=-\upDelta_{ba}(x,x')$, so $\upDelta$ is an anti-symmetric covector bi-distribution.
We remark that Lorentz boost invariance is broken because the crystal four-velocity $U$ and optic axis $X$ are preferred directions.

%++++++++++++++++++++++++++++++++++++++++++++++++++++++++++++++++++++++++++++++%
\subsection{Positivity and the microlocal spectrum condition}
\label{sub:uniaxial:microlocal}

In this subsection we show that the Pauli--Jordan propagator and positive frequency
two-point function meet the general conditions required to formulate the algebra~$\mathfrak{A}$ of smeared quantum fields and satisfy the microlocal spectrum condition.

Beginning with the Pauli--Jordan propagator, $\upDelta_{ab}(x,x')$ is evidently an antisymmetric bi-distribution.
It restricts to conserved compactly supported vector densities as a bilinear form
\begin{equation*}
  \sigma(j,j')\defn \upDelta(j,j') = \int_{\mathbb{R}^4\times\mathbb{R}^4} \upDelta_{ab}(x,x') j^a(x) j^{\prime b}(x') \,\dif x\, \dif x'.
\end{equation*}
with the property that $\sigma(j, j') = 0$ for all $j'$ if and only if $j = P A$ for some compactly supported $1$-form $A$.
This fact was already stated in \cite{Pfeifer:2016har}, however a precise proof was missing.
For completeness, we sketch the required argument here, which uses the fact that $E^{\text{ret/adv}}$ are inverses of $P$ up to gauge.
If $\sigma(j,j')=0$ for all $j'$, then $E^{\text{adv}}j$ and $E^{\text{ret}}j$ must be equal up to a pure gauge term.
Therefore their exterior derivatives are equal and, recalling the support properties of $E^{\text{ret/adv}}$, compactly supported.
As $F=\dif E^{\text{ret}} j$ is closed and compactly supported, there exists a compactly supported $1$-form~$A_0$ such that $F=\dif A_0$ and therefore $E^{\text{ret}} j = A_0 + \dif\chi$ for some smooth $\chi$.
Here we have used the Poincar\'e lemma in both the compact support and unrestricted forms~\cite{BottTu}.
It follows that $j=PE^{\text{ret}} j = PA_0$. Conversely, if $j=PA_0$ for some compactly supported $1$-form $A_0$, then $E^{\text{ret/adv}}j=E^{\text{ret/adv}}PA_0$ are equal up to a pure gauge term, and so $\sigma(j,j')=0$ for all $j'$.
Thus the Pauli-Jordan propagator defines the desired bilinear form required in Sect.~\ref{sub:pmedyn:quant} to define the commutator of the algebra of quantum fields.

Turning to $\upDelta^+$, \eqref{eq:gCCR} together with the expression $\Delta^-=(\Delta^+)^T$ derived above show that the generalized CCRs are fulfilled, \ie, $\upDelta^+-(\upDelta^+)^T=\im\upDelta$.
The microlocal spectrum condition requires the calculation of $\WF(\upDelta^+)$, which is most conveniently performed by using translational invariance to give
\begin{equation*}
  \WF(\upDelta^+)=\bigcup_{a,b=0}^3\WF(\upDelta_{ab}^+) = \bigcup_{a,b=0}^3 \big\{(x,k;x',-k)\in T^*\mathbb{R}^4\times T^*\mathbb{R}^4\;\big|\; (x-x',k)\in \WF(\tilde\upDelta_{ab}^+) \big\},
\end{equation*}
where the scalar distributions $\tilde\upDelta_{ab}^+$ are defined so that $\upDelta_{ab}^+(x,x')=\tilde\upDelta_{ab}^+(x-x')$.
To show that $\Delta^+$ obeys \eqref{eq:microlocal} it will suffice to show that $\WF(\tilde\upDelta_{ab}^+)\subset \mathcal{N}^+$.
Because $\mathcal{G}(k)\mathcal{E}_{ab}(k)$ is holomorphic, each $\tilde\upDelta_{ab}^+$ is a solution for the partial differential operator $\mathcal{G}(i\partial)$, and it follows that $\WF(\tilde\upDelta_{ab}^+)\subset \mathcal{N}$, which is the corresponding characteristic set.
We must therefore show that there are no directions from $\mathcal{N}^-$ in the wave-front set.
This can be seen from the computation
\begin{equation*}
  \widehat{f\tilde{\upDelta}_{ab}^+}(\ell)
  = \frac{\im}{(2\uppi)^4} \int_{\RR^3} \int_{C^+} \mathcal{E}_{ab}(k)\, \hat{f}(\ell-k)\, \dif k_0\, \dif \vec{k},
\end{equation*}
for any test function $f\in C_0^\infty(\mathbb{R})$.
As $k_0$ has nonnegative real part in the integration region, it is easily checked that the above expression decays rapidly as $\ell\to\infty$ in any cone within $(-\infty,0)\times\mathbb{R}^3$.
Every direction in $\mathcal{N}^-$ is therefore a regular direction for $\tilde\upDelta_{ab}^+$ and hence excluded from the wave-front set.

Next, we show that $\Delta^+$ obeys the hermiticity and positive type conditions when restricted to the space of conserved vector densties.
To do this, it is useful to decompose the tensors~$\mathcal{U}_{ab}$ and~$\tilde{\mathcal{U}}_{ab}$ as sums of manifestly positive rank-$1$ tensors and additional `pure gauge' terms containing either~$k_a$ or~$k_b$, which vanish when contracted with any vectors $V_1^a,V_2^a$ for which $k\cdot V_i=0$ ($i=1,2$).
Starting with $\tilde{\mathcal{U}}_{ab}$, an obvious possibility is to decompose $\tilde{\mathcal{U}}_{ab} = \tilde{u}_a \tilde{u}_b + \text{pure gauge}$, where
\begin{equation*}
  \tilde{u}_a \defn \frac{q_a}{\sqrt{(1+\xi^2)\eta^{-1}(q,q)}}\bigg|_{k_0 = \omega} = \frac{1}{\sqrt{k_2^2 + k_3^2}} \begin{pmatrix} k_1 \\ \tilde\omega \\ 0 \\ 0 \end{pmatrix}.
\end{equation*}
However, this covector diverges as $k_2^2+k_3^2\to 0$, \ie, in the limit where the extraordinary and ordinary lightcones touch.
Instead, we use a gauge-modified version
\begin{align}\label{eq:polVt}
  \tilde{v}_a(\vec{k}) \defn \tilde{u}_a(\vec{k}) - \frac{k_1 k_a}{\tilde{\omega}\sqrt{k_2^2+k_3^2}}
  &= \frac{1}{\tilde\omega\sqrt{k_2^2 + k_3^2}} \begin{pmatrix} 0 \\ (k_2^2+k_3^2)/(1+\xi^2) \\ -k_1 k_2 \\ -k_1 k_3 \end{pmatrix}
\end{align}
which satisfies $\zeta^{-1}(\tilde v,\tilde v)=1$ and remains bounded as $k_2^2+k_3^2\to 0$, giving a decomposition $\tilde{\mathcal{U}}_{ab} = \tilde{v}_a \tilde{v}_b + \text{pure gauge}$.

Turning to $\mathcal{U}_{ab}$, we note that the (non-zero) tensor
\begin{equation*}
  \mathcal{V}_{ab}(\vec{k}) \defn \eta_{ab} - \frac{q_a(k) q_b(k)}{\eta^{-1}(q,q)} \bigg|_{k_0 = \omega} + \frac{2 (k\cdot X) X_{(a}k_{b)}-\eta(X,X) (k_a k_b + 2 (k\cdot U) U_{(a}k_{b)})}{\eta^{-1}(q,q)}\bigg|_{k_0 = \omega}
\end{equation*}
annihilates $U^a$, $X^a$ and $k^a$ and therefore has has rank-$1$.
Indeed, one has $\mathcal{V}_{ab}=v_a v_b$, where
\begin{equation}\label{eq:polV}
  v_a(\vec{k}) \defn \frac{1}{\sqrt{k_2^2 + k_3^2}} \begin{pmatrix} 0 \\ 0 \\ k_3 \\ -k_2 \end{pmatrix}
\end{equation}
obeys the normalization condition $\eta^{-1}(v,v)=1$; as $\mathcal{V}_{ab}$ differs from $\mathcal{U}_{ab}$ only by pure gauge terms, we have ${\mathcal{U}}_{ab} = {v}_a {v}_b + \text{pure gauge}$.

Evidently the \emph{polarization covectors} $v$ and $\tilde v$ satisfy an analogue of the Coulomb gauge: writing $k_a=(\omega,\vec{k})$, $\tilde k_a = (\tilde\omega,\vec{k})$, we have
\begin{equation*}
  \tilde v \cdot U = v \cdot U = 0,
  \qquad\text{and}\qquad
  \eta^{-1}(k,v) = 0 = \zeta^{-1}(\tilde k, \tilde v)
\end{equation*}
for all $\vec{k}$.
Evaluating $\Delta^+(j,j')$, the pure gauge terms drop out and one has
\begin{equation*}
  \upDelta^+(j,j') = \frac{1}{(2\uppi)^3} \int_{\RR^3} \biggl( \frac{v_a(\vec{k})\what\jmath^{\,a}(-k)v_b(\vec{k})\what\jmath^{\,\prime b}(k)}{2\omega} + \frac{\tilde{v}_a(\vec{k})\what\jmath^{\,a}(-\tilde k)\tilde{v}_b(\vec{k})\what\jmath^{\,\prime b}(\tilde k)}{2\tilde\omega} \biggr)\,  \dif\vec{k}
\end{equation*}
for all conserved compactly supported vector densities $j,j'$.
Hermiticity holds because $v$ and $\tilde{v}$ are real, while the positivity condition is satisfied because
\begin{equation*}
  \upDelta^+(\conj{\jmath},j) = \frac{1}{(2\uppi)^3} \int_{\RR^3} \biggl( \frac{|v_a(\vec{k})\what\jmath^{\,a}(k)|^2}{2\omega} + \frac{|\tilde{v}_a(\vec{k})\what\jmath^{\,a}(\tilde k)|^2}{2\tilde\omega} \biggr)\, \dif\vec{k} \ge 0.
\end{equation*}

Summarizing, the positive frequency solution $\Delta^+$ obeys all the conditions required to define a physical quasi-free state $\Omega$ on the algebra $\mathfrak{A}$, completely determined by
\begin{equation}\label{eq:state}
  \Omega\bigl( \what{A}(j) \what{A}(j') \bigr) \defn \upDelta^+(j,j').
\end{equation}
Below, this will be shown to be a ground state with respect to time translations.
By construction, $\Delta^+$ extends the two-point function of $\Omega$ to a bi-distribution.

%++++++++++++++++++++++++++++++++++++++++++++++++++++++++++++++++++++++++++++++%
\subsection{Fock space and quantum fields}
\label{sub:uniaxial:qfield}

It will be useful to have a Hilbert space representation of $\mathfrak{A}$ available, in which the `vacuum' state~$\Omega$ defined by~\eqref{eq:state} is a vector state.
This can be done using the bosonic Fock space over a one-particle space
\begin{equation*}
  L^2(\mathbb{R}^3,\dif\vec{k}/(2\uppi)^3) \otimes \CC^2.
\end{equation*}
In familiar notation, this Fock space carries a quantum field $\what{A}_a(x) = \what{A}_a(t, \vec{x})$ given by
\begin{equation}\label{eq:qfield}\begin{split}
  \what{A}_a(t, \vec{x})
  \defn \frac{1}{(2\uppi)^3} \int_{\RR^3} &\biggl( a(\vec{k}) \frac{v_a(\vec{k})}{\sqrt{2\omega}} \e^{-\im(\vec{k} \cdot \vec{x} + \omega t)}
  + \tilde{a}(\vec{k}) \frac{\tilde{v}_a(\vec{k})}{\sqrt{2\tilde\omega}} \e^{-\im(\vec{k} \cdot \vec{x} + \tilde\omega t)} + \text{h.c.} \biggr)\, \dif\vec{k},
\end{split}\end{equation}
where the annihilation and creation operators obey the CCR
\begin{equation*}
  \bigl[ a(\vec{k}), a^*(\vec{k}') \bigr]
  = \bigl[ \tilde{a}(\vec{k}), \tilde{a}^*(\vec{k}') \bigr]
  = (2\uppi)^3 \delta^{(3)}(\vec{k}-\vec{k}') \one
\end{equation*}
with all other commutators vanishing.
The integral in~\eqref{eq:qfield} includes rays along the optic axis, at which the polarization covectors $v$ and $\tilde{v}$ have discontinuous, direction-dependent, limits.
As they remain bounded, however, \eqref{eq:qfield} is well-defined; what is required is that maps such as $j\mapsto (2\omega)^{-1/2}v\cdot\what\jmath|_{k_0=\omega}$ are well-defined maps from test vector densities to $L^2(\mathbb{R}^3,\dif\vec{k}/(2\uppi)^3)$.

Smearings of $\what{A}$ against conserved vector densities, and sums of products thereof, provide a representation of $\mathfrak{A}$.
For example, it is not difficult to verify directly that $\what{A}_a$ solves the field equations~\eqref{eq:A-field-eq}.
Indeed, the computation reduces to the verification of
\begin{equation*}
  \chi^{acbd} v_b k_c k_d = 0,
  \quad
  \chi^{acbd} \tilde{v}_b \tilde{k}_c \tilde{k}_d = 0
\end{equation*}
on the ordinary resp. extraordinary lightcone, easily proved using~\eqref{eq:constitutive}, \eqref{eq:polV} and \eqref{eq:polVt}.
The CCRs hold as a result of \eqref{eq:spatial-PJ} and the equality of $\mathcal{U}_{ab}$ and $v_a v_b$ (resp., $\tilde{\mathcal{U}}_{ab}$ and $\tilde v_a \tilde v_b$) up to pure gauge terms; in a similar way, one may compute directly that
\begin{equation*}
  \ip{\Omega}{\what{A}(j) \what{A}(j')\, \Omega} = \upDelta^+(j,j')
\end{equation*}
holds for conserved $j,j'$, where we have written $\Omega$ to denote also the Fock vacuum vector, annihilated by all $a(\vec{k})$ and $\tilde a(\vec{k})$.
Note that $\what{A}$ can be smeared against any smooth compactly supported vector density to give a Hilbert space operator, but only smearings against conserved vector densities yield operators representing elements of $\mathfrak{A}$.

Although boost invariance is broken in the crystal background, translational invariance is maintained.
In particular, time translations are generated by the Hamiltonian
\begin{equation*}
  H = \frac{1}{(2\uppi)^3} \int_{\RR^3} \bigl( \omega a^*(\vec{k}) a(\vec{k}) + \tilde\omega \tilde{a}^*(\vec{k}) \tilde{a}(\vec{k}) \bigr)\, \dif\vec{k}
\end{equation*}
with respect to which $\Omega$ is clearly a ground state.

Starting from these definitions, one can introduce operators corresponding to other observables.
For example, the quantized field strength
\begin{equation*}\begin{split}
  \what{F}_{ab}(x)
  = -\frac{2\im}{(2\uppi)^3} \int_{\RR^3} \biggl(& a(\vec{k}) \frac{k_{[a} v_{b]}(\vec{k})}{\sqrt{2\omega}} \e^{-\im(\vec{k} \cdot \vec{x} + \omega t)}
  +\tilde{a}(\vec{k}) \frac{\tilde{k}_{[a} \tilde{v}_{b]}(\vec{k})}{\sqrt{2\tilde\omega}} \e^{-\im(\vec{k} \cdot \vec{x} + \tilde\omega t)} - \text{h.c.} \biggr)\, \dif\vec{k}
\end{split}\end{equation*}
can be directly obtained from~\eqref{eq:qfield}.

%******************************************************************************%
\section{QEI for the uniaxial crystal}
\label{sec:uniaxial-qei}

In this section, we first demonstrate the existence of negative energy density states in the uniaxial crystal, then show that the QEI derived in Section~\ref{sec:qei} holds for subluminal trajectories and evaluate the bound explicitly. Among other things this involves an explicit proof that the classical sWEC holds on subluminal trajectories. The situation is different for interluminal trajectories: the classical sWEC fails and, consequently, so do the QEI bounds. A subtle point is also addressed: in the pre-metric situation there is no preferred proper time normalization of observer trajectories. Accordingly, we discuss normalizations arising both from the background Minkowski metric $\eta$ and intrinsically generated from the pre-metric theory, and trace the effect on our results.

%++++++++++++++++++++++++++++++++++++++++++++++++++++++++++++++++++++++++++++++%
\subsection{States with locally negative energy density}
\label{sub:uniaxial-qei:neg}

In the Fock space, normal ordering with respect to the state~$\Omega$ can be achieved by the standard normal ordering of annihilation and creation operators.
Computing the normal ordered energy density operator $\norder{\what\rho(f)}$ in this way for a given choice of frame, we adapt a simple argument here to demonstrate that there exist states with locally negative energy density expectation values.
Consider the quantum states defined by the family of vectors
\begin{equation*}
	\Psi(\phi) \defn \cos\phi\ \Omega + \sin\phi\, \norder{\what\rho(f)} \Omega,
  \quad
  \phi \in \left[ -\tfrac\uppi2,\tfrac\uppi2 \right],
\end{equation*}
where $\Omega$ is the Fock vacuum vector discussed in Sect.~\ref{sub:uniaxial:qfield} and $f\in C_0^\infty(\RR^4)$ is a real-valued test function, normalized so that $\norm{\norder{\what\rho(f)} \Omega} = 1$.
(As shown in Appendix~\ref{appx:ReehSchlieder}, one can exclude the possibility that $\norder{\what\rho(f)} \Omega=0$.)
Calculating the expectation value of the quantized energy density in the state given by $\Psi(\phi)$ yields
\begin{align*}
	\langle \norder{\what\rho(f)} \rangle_{\Psi(\phi)}
	&= \sin(2\phi) \norm{\norder{\what\rho(f)} \Omega}^2 + \sin^2\phi\, \ip{\Omega}{\norder{\what\rho(f)}^3\, \Omega} \\
	&= 2\phi + \mathcal{O}(\phi^2).
\end{align*}
Choosing $-\phi$ sufficiently small, we thus see that there exist states such that the expectation value of the quantized energy density becomes negative and therefore the pointwise energy density must also be negative on an open set within the support of $f$.
By translation, one can arrange that this occurs in any desired region of any given worldline of interest.
Owing to the quantum energy inequality, however, the expectation of $\norder{\what\rho(f)}$ cannot become arbitrarily negative in the states $\Psi(\phi)$ or any other states satisfying the microlocal spectrum condition.

%++++++++++++++++++++++++++++++++++++++++++++++++++++++++++++++++++++++++++++++%
\subsection{Quantized point-split energy density}
\label{sub:uniaxial-qei:energy}

We now begin the explicit computation of the QEI along trajectories with uniform velocity relative to the crystal.
The first step is to obtain the point-split energy density, for which purpose we may use~\eqref{eq:psplitconst} instead of~\eqref{eq:classPS} due to translational invariance.
However, the calculations that would be needed to use~\eqref{eq:classPS} can be read off from Sect.~\ref{sub:uniaxial-qei:swec}.
The QEI bound itself will be derived in Sect.~\ref{sub:uniaxial-qei:subluminal}.

%..............................................................................%
\subsubsection{General expression}

Let $\gamma$ be a subluminal trajectory, equipped with a frame $e$ such that $\varepsilon(e)=|\eta|^{1/2}$ (equal to unity in the global Cartesian coordinates).
Evaluating the point-split energy density~\eqref{eq:psplitconst} in the state defined by the two point function~\eqref{eq:posneg_freq}, we obtain
\begin{equation*}\begin{split}
	\rho_\Omega(\tau,\tau') &= \frac{1}{2(2\uppi)^3} (\chi^{abcd} - 2 \chi^{ebcd} n_e \dot\gamma^a - 2 \chi^{abed} n_e \dot\gamma^c) \\&\quad \times \int_{\RR^3} \biggl( k_{[a} \mathcal{U}_{b][d} k_{c]} \frac{\e^{-\im k(\gamma(\tau) - \gamma(\tau'))}}{2 \omega} + \tilde{k}_{[a} \tilde{\mathcal{U}}_{b][d} \tilde{k}_{c]} \frac{\e^{-\im \tilde{k}(\gamma(\tau) - \gamma(\tau'))}}{2\tilde\omega} \biggr)\, \dif\vec{k},
\end{split}\end{equation*}
where we again use the notation $k = (\omega,\vec{k})$ and $\tilde{k} = (\tilde\omega,\vec{k})$ for the ordinary and extraordinary null covectors given by $\vec{k}$.

Specializing to the case of a constant velocity inertial observer $\gamma(\tau) = \dot{\gamma} \tau$ (maintaining the condition $\varepsilon(e)=|\eta|^{1/2}$), we employ the definition of~$\mathcal{U}$ and~$\tilde{\mathcal{U}}$ in~\eqref{eq:Uab1} and~\eqref{eq:Uab2} to obtain from a straightforward but lengthy calculation
\begin{equation}\label{eq:LambdaRhoUniaxial}\begin{split}
  \rho_\Omega(\tau,\tau') = -\frac{1}{2(2\uppi)^3} \int_{\RR^3} &\biggl( \frac{(k \cdot \dot\gamma)\, \eta^{-1}(n,k)}{\omega}\, \e^{-\im k\cdot\dot{\gamma}(\tau-\tau')} \\&\phantom{\biggl(}\mathllap{+\mskip\medmuskip} \frac{(\tilde{k} \cdot \dot\gamma)\, \zeta^{-1}(n,\tilde{k})}{(1+\xi^2) \tilde\omega}\, \e^{-\im \tilde{k} \cdot\dot{\gamma}(\tau-\tau')} \biggr)\, \dif\vec{k},
\end{split}\end{equation}
with Minkowski spacetime limit $\xi \to 0$
\begin{equation*}
  \rho_\Omega(\tau,\tau') = -\frac{1}{(2\uppi)^3} \int_{\RR^3} \frac{(k \cdot \dot{\gamma})\, \eta^{-1}(n,k)}{\omega}\, \e^{-\im k \cdot\dot{\gamma}(\tau - \tau')}\, \dif\vec{k}.
\end{equation*}
To gain more insight about this expression we will evaluate it more explicitly for sub- and interluminal trajectories.

%..............................................................................%
\subsubsection{Expression for subluminal trajectories}
\label{ssub:uniaxial-qei:energy:subluminal}

Consider a uniform velocity trajectory that is $\eta$-timelike, and therefore is either subluminal or interluminal.
Let $\alpha\in \RR$ be its rapidity, in the rest frame of the crystal, and $\beta\in (-\uppi,\uppi]$ be the angle made between the $3$-velocity and the positive $x$-axis, \ie, the optic axis.
Without loss of generality (rotating the coordinate system in the $yz$-plane if necessary) the worldline takes the form
\begin{equation}\label{eq:worldline}
  \gamma(\tau) = \N\, \tau\, (\cosh\alpha, \sinh\alpha \cos\beta, 0, \sinh\alpha \sin\beta).
\end{equation}
Then $\dot\gamma = \N\, (\cosh\alpha, \sinh\alpha \cos\beta, 0, \sinh\alpha \sin\beta)$ is constant and this vector may be extended to a frame $e$ with $e_0=\dot\gamma$ and $\varepsilon(e)=|\eta|^{1/2}$, and so that the dual basis covector $n=e^{*0}$ is
\begin{equation*}
  n = \N^{-1}\, (\cosh\alpha, -\sinh\alpha \cos\beta, 0, -\sinh\alpha \sin\beta).
\end{equation*}
The normalization factor $\N$ in~\eqref{eq:worldline} was introduced to trace how our results~\eqref{eq:pointsplit} and~\eqref{eq:qeiuni} depend on the parametrization of the worldlines, and will be discussed in Sect.~\ref{sub:uniaxial:normalization}.

While the worldline~$\gamma$ is $\eta$-timelike by construction, it is not necessarily timelike with respect to~$\zeta$ because
\begin{align*}
  \zeta(\dot\gamma, \dot\gamma)
  % &= \eta(\dot\gamma,\dot\gamma) + \frac{\xi^2}{1+\xi^2}\eta(U,\dot\gamma)^2 - \frac{1}{1+\xi^2}\eta(X,\dot\gamma)^2\\
  &= \N^2\, \frac{\xi^2\sinh^2\alpha\sin^2\beta-1}{1+\xi^2},
\end{align*}
see~\eqref{eq:zeta}.
Therefore the trajectory is subluminal if
\begin{align*}
  \sinh^2\alpha\sin^2\beta &< \xi^{-2}
\shortintertext{and interluminal if}
  \sinh^2\alpha\sin^2\beta &>\xi^{-2}.
\end{align*}
This distinction has another significance: as we show in Sect.~\ref{sub:uniaxial-qei:swec}, the sWEC holds for subluminal trajectories but fails in the interluminal case.

For the rest of this subsection, and also subsection~\ref{sub:uniaxial-qei:subluminal}, we will assume that $\gamma$ is subluminal, so $\dot{\gamma}$ is timelike with respect to both metrics~$\eta$ and~$\zeta$.
In this case, the integrals in~\eqref{eq:LambdaRhoUniaxial} may be calculated using identities~\eqref{eq:abk} and~\eqref{eq:abkt} proved in Appendix~\ref{appx:int} to give
\begin{equation}\label{eq:pointsplit}
  \rho_\Omega(\tau,\tau') = \frac{C(\alpha,\beta,\xi)}{(2\uppi)^2\, \N^4} \int_0^\infty \kappa^3 \e^{-\im\kappa (\tau-\tau')}\, \dif\kappa,
\end{equation}
where
\begin{align}
C(\alpha,\beta,\xi) &\defn \N^4
\left(\frac{n\cdot\dot{\gamma}}{\eta(\dot{\gamma},\dot{\gamma})^2} + \frac{n\cdot\dot{\gamma}}{\zeta(\dot{\gamma},\dot{\gamma})^2}
  \right) \nonumber \\
  &\mathrel{\phantom{\defn}\mathllap{=}} 1 + (1+\xi^2) \bigl(1 - \xi^2 \sinh^2\alpha \sin^2\beta\bigr)^{-2} \nonumber \\
  &\mathrel{\phantom{\defn}\mathllap{=}} 2 +  (1+2\sinh^2\alpha\sin^2\beta) \xi^2+\mathcal{O}(\xi^4). \label{eq:pointsplitNsrt}
\end{align}

\begin{figure}
  \begin{minipage}[t]{.31\textwidth}\centering
    \includegraphics[width=\textwidth]{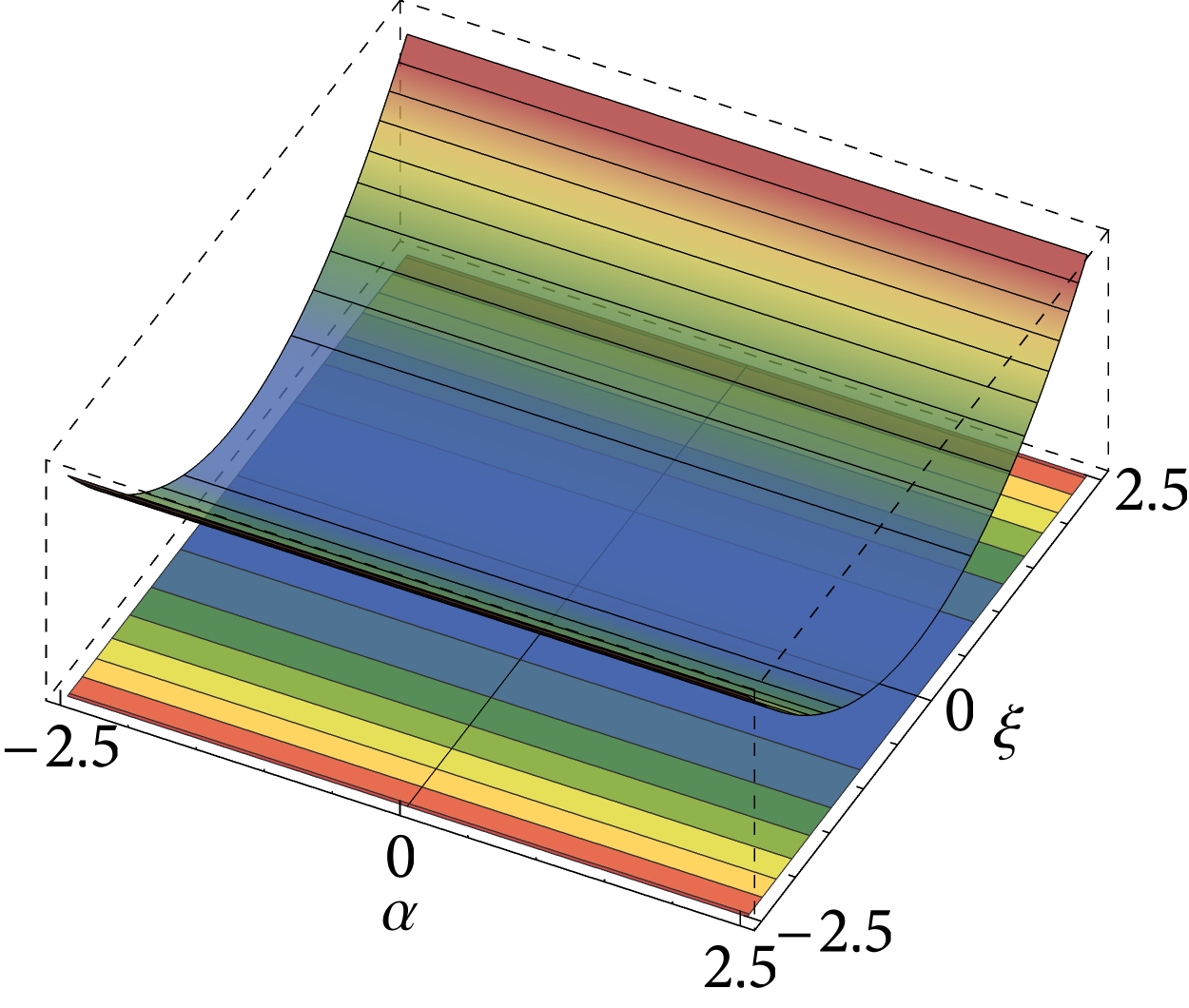}
  \end{minipage}
  \hfill
  \begin{minipage}[t]{.31\textwidth}\centering
    \includegraphics[width=\textwidth]{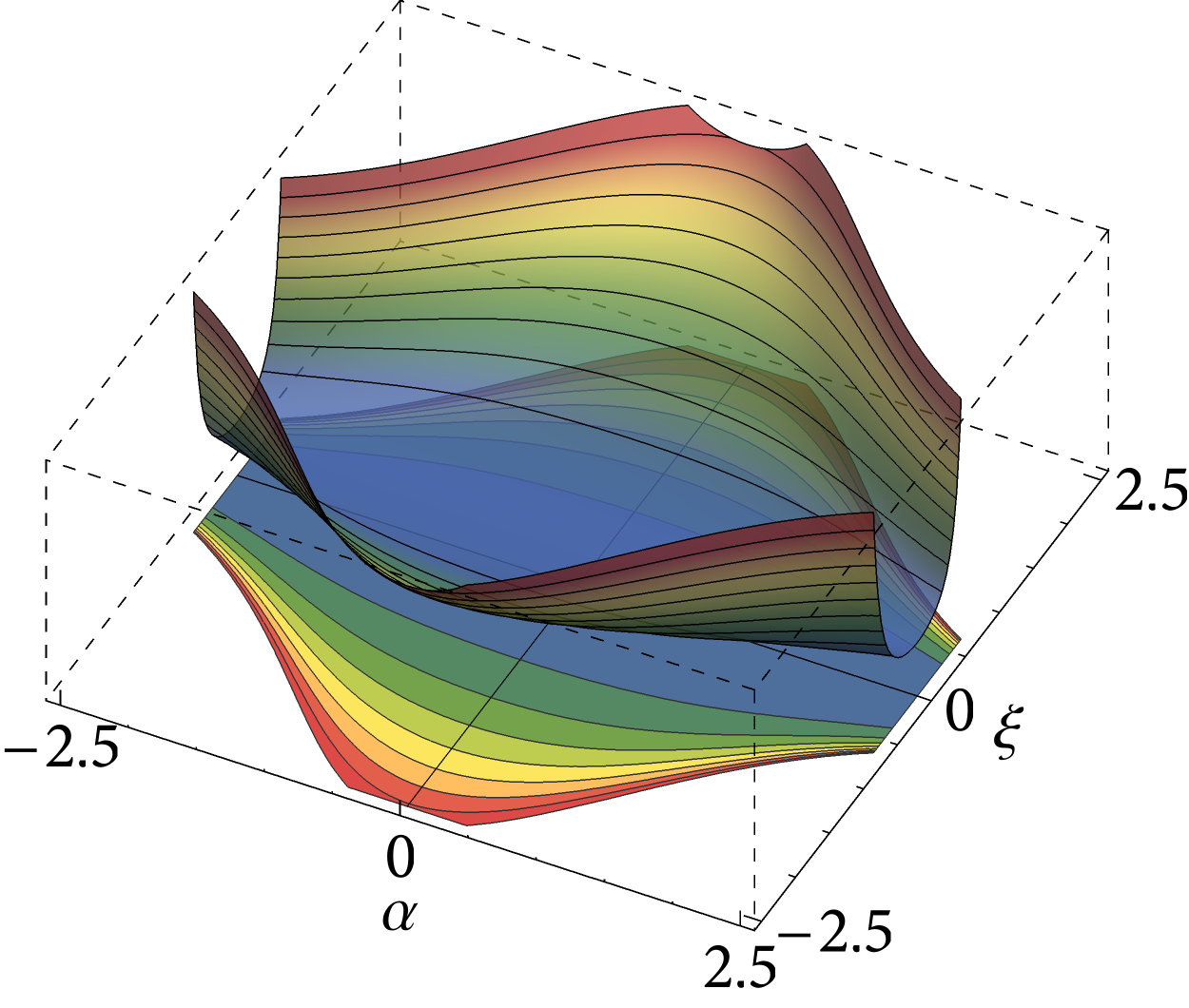}
  \end{minipage}
  \hfill
  \begin{minipage}[t]{.31\textwidth}\centering
    \includegraphics[width=\textwidth]{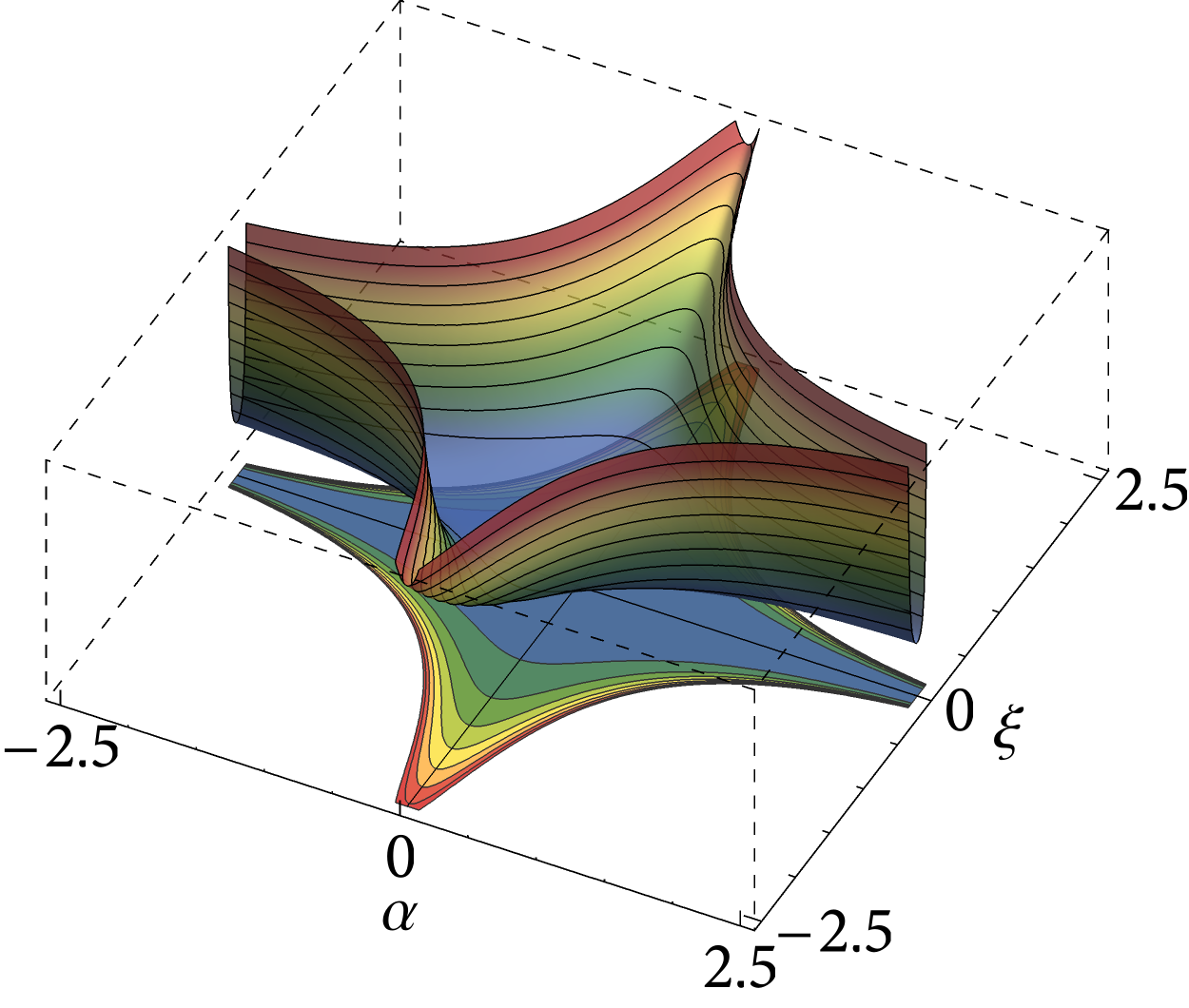}
  \end{minipage}
  \caption{Plot of $C(\alpha,\beta,\xi)$ with $\alpha$ and $\xi$ in the range $[-2.5,2.5]$, $\beta = 0, \frac{\uppi}{16}, \frac{\uppi}{2}$ (from left to right), and $C(\alpha,\beta,\xi)$ cut off at the value $9$.}
  \label{fig:plot}
\end{figure}

Note that $C(\alpha,\beta,\xi) \to 2$ as $\xi \to 0$ with $\alpha,\beta$ fixed, which reproduces the known Lorentz-invariant Minkowski spacetime result for electromagnetism~\cite{fewster:2003}.
On the other hand, fixing $\xi$ we see that $C(\alpha,\beta,\xi) \to 2 + \xi^2$ as $\alpha \to 0$ or $\beta\to 0$, but $C(\alpha,\beta,\xi) \to +\infty$ as $\sinh \alpha \sin\beta\to \pm\xi^{-1}$, \ie, $C(\alpha,\beta,\xi)$ diverges for worldlines which become lightlike with respect to the extraordinary lightcone given by $\zeta$.
The shape of $C(\alpha,\beta,\xi)$ can be seen in Fig.~\ref{fig:plot} in the cases $\beta=0,\frac\uppi{16},\frac\uppi2$.

As expected the result \eqref{eq:pointsplit} depends on the normalization $\N$ of the worldline under consideration, whose influence we will discuss now.

%..............................................................................%
\subsubsection{Normalization factor}
\label{sub:uniaxial:normalization}

There are different viewpoints from which one can interpret this uniaxial crystal electrodynamics.
One may regard it as describing a crystal in Minkowski spacetime which is probed by observers whose dynamics are determined by special relativity.
Alternatively, one may see the constitutive law of the crystal as fundamental and demand that probes obey the point particle dynamics and causal behavior dictated from the theory of electrodynamics which it defines.
For our purposes, the difference between the two viewpoints lies in the way how one parametrizes the trajectory of the probe, along which one calculates the QEI.
The choice of parametrization is reflected in the quantized point-split energy density \eqref{eq:pointsplit} by the appearance of normalization factor $\N$.

In the context of special (SR) and general relativity timelike worldlines $\gamma$ are observer worldlines if they are proper time parametrized, \ie, satisfy $\eta(\dot\gamma, \dot\gamma)=-1$ resp. $\eta$ replaced by a general spacetime metric.
For the curves we are considering this corresponds to the choice $\N=\N_{\mathrm{SR}}=1$, which represents observers not influenced by the crystal background structure.

An alternative choice for $\N$ is to consider observers subject to the dispersion relation induced by electrodynamics.
The Fresnel polynomial of the uniaxial crystal (UC) defines a natural massive dispersion relation $|\eta|^{-1/2}\mathcal{G}(k)=m^4$ and this in turn gives a natural parameterization of the motion of probe particles with mass $m$ \ie\ their proper time.
The precise mathematical methods to derive this normalization were discussed in detail in~\cite{Raetzel:2010je}.
In this approach the normalization $\N = \N_{\mathrm{UC}}$ depends on the crystal parameter $\xi$, the rapidity $\alpha$ and the angle $\beta$ made with the optic axis, modifying the point-split energy density along the curve.
As a matter of fact the calculations needed to compute the normalization as a function $\N(\alpha,\beta,\xi)$ for massive point particles governed by this electrodynamically induced clock are non-trivial, which is why we only derive it to third order in the crystal parameter
\begin{equation*}
  \N_{\mathrm{UC}} = 1 - \frac{\xi^2}{4}(1 + \sinh^2\alpha \sin^2\beta) + \mathcal{O}(\xi^4).
\end{equation*}
The derivation can be found in Appx.~\ref{appx:MassiveDual}.
The all order calculation is beyond the scope of this work and may be investigated in the future.
What we can be deduced to all orders already at this point is that $\N_\xi(\alpha,\beta) = \N(\alpha,\beta,\xi)$ is a smooth non-vanishing function of the rapidity parameter $\alpha$ and the angle $\beta$.
This behavior is guaranteed by the smoothness and invertibility properties of the Legendre map (see Sect. \ref{sub:pmedyn:obs} around \eqref{eq:Legendre} and \cite{Raetzel:2010je}) as a map from massive momenta, \ie, the interior of the hyperbolicity cone $\Gamma$ of the dispersion relation, onto velocities inside the dual $\Gamma^+$ of the hyperbolicity cone.

For the dependence of the point-split energy density \eqref{eq:pointsplit} on the rapidity $\alpha$, the angle $\beta$ and the crystal parameter $\xi$ this means that for the special relativistic observer normalization $\N_{\mathrm{SR}}=1$, $C(\alpha,\beta,\xi)$ determines the behaviour of the point-split energy density completely.
For the alternative normalization $\N_{\mathrm{UC}}$ to second order in the crystal parameter we find the modified dependence
\begin{equation}\label{eq:pointsplitNuc}
  \frac{C(\alpha,\beta,\xi)}{\N_{\mathrm{UC}}^4} = 2+(3+4\sinh^2\alpha\sin^2\beta)\xi^2 +\mathcal{O}(\xi^4),
\end{equation}
which nicely illustrates the influence of the normalization factor explicitly by comparing \eqref{eq:pointsplitNsrt} and \eqref{eq:pointsplitNuc}.

%++++++++++++++++++++++++++++++++++++++++++++++++++++++++++++++++++++++++++++++%
\subsection{Proof of the sWEC}
\label{sub:uniaxial-qei:swec}

To show that the general quantum energy inequality derived in Sect.~\ref{sec:qei} holds for subluminal trajectories in the uniaxial crystal considered here, we also need to show that the strict form of the weak energy condition (sWEC) holds for such trajectories.
We will show this using the decomposition of $\rho$ into $\chi_1$ and $\chi_2$ in~\eqref{eq:chi_12-decomp}, acting on the subspace of `magnetic' and `electric' $2$-forms, respectively.
More precisely, we will use the formulas~\eqref{eq:chi_12-6x6_repr} to express $\chi_1$ and $\chi_2$ as $3 \times 3$ matrices in a judiciously chosen basis.

Choose the following dual frame for the magnetic and electric subspaces
\begin{subequations}\label{eq:subspace_basis}\begin{align}
  [\mathfrak{b}_1^{ab}] &= \bigl( 0, -\sinh\alpha \sin\beta, 0, 1 + (\cosh\alpha-1) \sin^2\beta, 0, (1-\cosh\alpha) \cos\beta \sin\beta \bigr), \\
  [\mathfrak{b}_2^{ab}] &= \bigl( \sinh\alpha \sin\beta, 0, -\sinh\alpha \cos\beta, 0, \cosh\alpha, 0 \bigr), \\
  [\mathfrak{b}_3^{ab}] &= \bigl( 0, \sinh\alpha \cos\beta, 0, (1-\cosh\alpha) \cos\beta \sin\beta, 0, 1 + (\cosh\alpha-1) \cos^2\beta \bigr), \\
  [\mathfrak{e}_1^{ab}] &= C^{-\frac12} \bigl( 1 + \sinh^2\alpha \sin^2\beta, 0, -\sinh^2\alpha \cos\beta \sin\beta, 0, \cosh\alpha \sinh\alpha \sin\beta, 0 \bigr), \\
  [\mathfrak{e}_2^{ab}] &= \bigl( 0, \cosh\alpha, 0, -\sinh\alpha \sin\beta, 0, \sinh\alpha \cos\beta \bigr), \\
  [\mathfrak{e}_3^{ab}] &= C^{-\frac12} \bigl( 0, 0, -\cosh\alpha, 0, \sinh\alpha \cos\beta, 0 \bigr),
\end{align}\end{subequations}
where $C = 1 + \sinh^2\alpha \sin^2\beta$ and the 6 entries give the $01, 02, 03, 23, 31, 12$ components, \ie,
\begin{align*}
  [\mathfrak{b}^{ab}_A] = (\mathfrak{b}^{01}_A, \mathfrak{b}^{02}_A, \mathfrak{b}^{03}_A, \mathfrak{b}^{23}_A, \mathfrak{b}^{31}_A, \mathfrak{b}^{12}_A)
\end{align*}
and similar for $\mathfrak{e}^{ab}_A$.
In this basis, we calculate (with the help of the computer algebra system \textit{Mathematica}\texttrademark)
\begin{align*}
  X_1^{AB} &= \begin{pmatrix}
    1 & 0 & 0 \\
    0 & 1 - \xi^2 \sinh^2\alpha \sin^2\beta & 0 \\
    0 & 0 & 1
  \end{pmatrix}, \\
  X_2^{AB} &= \begin{pmatrix}
    1 + \xi^2 (1 + \sinh^2\alpha \sin^2\beta) & 0 & 0 \\
    0 & 1 & 0 \\
    0 & 0 & 1
  \end{pmatrix}
\end{align*}
so that $\chi_1^{abcd} = X_1^{AB} \mathfrak{b}_1^{ab} \mathfrak{b}_1^{cd}$ and $\chi_2^{abcd} = X_2^{AB} \mathfrak{e}_1^{ab} \mathfrak{e}_1^{cd}$.
We immediately see that $X_2$ is always positive definite, while $X_1$ is positive definite if and only if $\sinh^2\alpha \sin^2\beta < \xi^{-2}$, which is exactly the condition of Sect.~\ref{ssub:uniaxial-qei:energy:subluminal} that the worldline $\gamma$ is subluminal.
This proves that sWEC holds inside the uniaxial crystal along all subluminal trajectories.
Simultaneously, this shows that the QEI derived in Sect.~\ref{sec:qei} holds.

Since the matrices $X_1, X_2$ are diagonal, we can easily take their square roots in the subluminal case:
\begin{align*}
  Y_1^{AB} &= \begin{pmatrix}
    1 & 0 & 0 \\
    0 & \sqrt{1 - \xi^2 \sinh^2\alpha \sin^2\beta} & 0 \\
    0 & 0 & 1
  \end{pmatrix}, \\
  Y_2^{AB} &= \begin{pmatrix}
    \sqrt{1 + \xi^2 (1 + \sinh^2\alpha \sin^2\beta)} & 0 & 0 \\
    0 & 1 & 0 \\
    0 & 0 & 1
  \end{pmatrix}.
\end{align*}
We could use this result to explicitly determine the QEI bound, \ie, the right-hand side of~\eqref{eq:QEI}.
However, due to translational invariance it is easier to use simpler methods as applied in the following section.

%++++++++++++++++++++++++++++++++++++++++++++++++++++++++++++++++++++++++++++++%
\subsection{QEI bound for subluminal trajectories}
\label{sub:uniaxial-qei:subluminal}

We can now give the explicit form of the quantum energy inequality~\eqref{eq:QEI} for the uniaxial crystal for curves which propagate slower than the extraordinary speed of light.
The statement of the QEI is that for all subluminal curves~\eqref{eq:worldline}, that is, for $\sinh^2\alpha \sin\beta< \xi^{-2}$, the normal-ordered energy density obeys
\begin{equation*}
  \int_\RR \abs{g(\tau)}^2 \langle \norder{\rho(\tau)} \rangle_\Lambda\,  \dif\tau
  \geq -\frac{1}{\uppi} \int_0^\infty \left( \iint_{\RR^2} g(\tau) g(\tau') \rho_\Omega(\tau,\tau')\, \e^{-\im \beta(\tau-\tau')}\, \dif\tau\, \dif\tau' \right) \dif\beta
\end{equation*}
for all states $\Lambda$ obeying the microlocal spectrum condition and all real valued compactly supported~$g$. Inserting the point-split energy density~\eqref{eq:pointsplit} and evaluating the resulting integrals by rearranging the order of integration and the Plancherel theorem, this becomes
\begin{align}\label{eq:qeiuni}
  \int_\RR \abs{g(\tau)}^2 \langle \norder{\rho(\tau)} \rangle_\Lambda\,  \dif\tau
  \geq
  &= -\frac{C(\alpha,\beta,\xi)}{\uppi (2\uppi)^2\, \N^4} \int_0^\infty \left( \int_0^\infty \kappa^3 \abs{\what{g}(\kappa+\beta)}^2\, \dif\kappa \right) \dif\beta \nonumber\\
  &= -\frac{C(\alpha,\beta,\xi)}{\uppi (2\uppi)^2\, \N^4} \int_0^\infty \left(\abs{\what{g}(\theta)}^2 \int_0^\theta \kappa^3\, \dif\kappa\right) \dif\theta \nonumber\\
  &= -\frac{C(\alpha,\beta,\xi)}{2(2\uppi)^3\, \N^4} \int_0^\infty \theta^4 \abs{\what{g}(\theta)}^2\, \dif\theta \nonumber\\
  &= -\frac{C(\alpha,\beta,\xi)}{4(2\uppi)^2\, \N^4} \norm*{g''}_2^2,
\end{align}
and the overall result can be extended to complex-valued test functions by applying the real result to the real and imaginary parts of $g$ separately.
Thus along subluminal trajectories there exists a finite negative bound on the quantized energy density of the electromagnetic field.
In the Minkowski spacetime limit $\xi\to0$ the bound becomes independent of $\alpha,\beta$ for either of the two normalizations discussed, since $C(\alpha,\beta,0)=2$ and $\N(\alpha, \beta,0)=1$.

For the observer at rest with respect to the crystal ($\alpha=0$) we find
\begin{equation*}
  \int_\RR \abs{g(\tau)}^2 \langle \norder{\rho(\tau)} \rangle_\Lambda\, \dif\tau \geq -\frac{2+\xi^2}{16\uppi^2\, \N^4} \norm*{g''}_2^2
\end{equation*}
and indeed this holds for any $\alpha$ if $\beta=0$, \ie, for motion along the optic axis.

However, for $\beta\neq 0$, the closer the observer's velocity comes to the extraordinary speed of light, \ie, the lightrays propagating along the cone of $\zeta$, the more negative the lower bound~\eqref{eq:qeiuni} becomes, diverging in the limits $\sinh\alpha \sin\beta\to\pm \xi^{-1}$.
Again this holds for either choice of normalizations discussed.

%++++++++++++++++++++++++++++++++++++++++++++++++++++++++++++++++++++++++++++++%
\subsection{Failure of QEIs along interluminal trajectories}
\label{sub:uniaxial-qei:interluminal}

Our QEI above was proved for averaging along subluminal trajectories, for which the classical sWEC holds as discussed in Sect.~\ref{sub:uniaxial-qei:swec}.
Here, we show that no QEI can hold along an interluminal trajectory ($\eta$-timelike and $\zeta$-spacelike) in the translationally invariant uniaxial birefringent crystal.
The argument is based on one introduced in~\cite{FewsterOsterbrink:2008} for non-minimally coupled scalar fields and shows that a failure of the classical sWEC for a positive energy solution entails a corresponding failure in the QEI.
It is valid for any constant velocity curve passing through $x=0$, and is thus independent of the parametrization and normalization of the curve.

The starting-point is the fact that single-particle states in QFT correspond to classical positive frequency solutions.
Let $\Psi$ be a vector state in the Fock space of the form
\begin{equation*}
  \Psi = \int_{\RR^3} \sqrt{2\tilde{\omega}(k_2^2 + k_3^2)}\, f(\vec{k})\, \tilde{a}^*(\vec{k})\, \Omega\, \dif\vec{k},
\end{equation*}
where $\Omega$ is the vacuum vector and $f\in\mathscr{S}(\RR^3)$ is a Schwartz function, chosen so that $\Psi$ is normalized.
(The factors in the square root are inserted for later convenience.)
Then, using the explicit form of the quantum field~\eqref{eq:qfield}, the corresponding positive frequency solution is
\begin{equation*}
  A_a(x) \defn \ip{\Omega}{\what{A}_a(x)\, \Psi}  = \int_{\RR^3} \sqrt{k_2^2 + k_3^2}\, f(\vec{k})\, \tilde{v}_a(\vec{k})\, \e^{-\im(\vec{k}\cdot\vec{x} + \tilde\omega t)}\, \dif\vec{k}
\end{equation*}
and is easily seen to be smooth as a consequence of the rapid decay of $f$.

As a consequence of Wick's theorem, the $n$-particle state $\Psi^{\otimes n}$ has two-point function
\begin{align*}
  \ip{\Psi^{\otimes n}}{\what{A}_a(x)\what{A}_b(y)\, \Psi^{\otimes n}} &=
  n \bigl( \overline{A_a(x)} A_b(y) + \overline{A_b(y)} A_a(x) \bigr) +
  \ip{\Omega}{\what{A}_a(x)\what{A}_b(y)\, \Omega}
\end{align*}
and therefore the normal ordered two-point function is
\begin{equation*}
  \ip{\Psi^{\otimes n}}{\norder{\what{A}_a(x)\what{A}_b(y)}\, \Psi^{\otimes n}} = 2n\Re \bigl(\overline{A_a(x)} A_b(y)\bigr).
\end{equation*}
It follows that the quantized energy density (defined with respect to any curve $\gamma$ and frame $e$) in the state $\Psi^{\otimes n}$ is $n$-times that in state $\Psi$, which in turn equals twice the corresponding complexified \emph{classical} energy density of the complex-valued solution $A_a(x)$ defined by
\begin{equation*}
  \rho = \frac18 \varepsilon(e)^{-1} \chi^{abcd} \bigl(\conj{F}_{ab} F_{cd} - 4 n_a \dot\gamma^e \Re(\conj{F}_{eb} F_{cd}) \bigr),
\end{equation*}
with $F=\dif A$ as usual.
One sees immediately that if $\rho<0$ at some point along $\gamma$ -- a failure of the classical sWEC for positive energy solutions -- then the quantum field theory cannot obey a QEI: any weighted average of the quantized energy density $\norder{\what{\rho}}$ along $\gamma$, supported in the region where the sWEC fails, has
a negative expectation value in the state $\Psi$, and hence its expectation value in state $\Psi^{\otimes n}$ is unbounded from below as $n\to\infty$.

It remains to show that $f$ may be chosen to violate the sWEC for a constant-velocity interluminal observer.
The solution is in the form of a wave packet of extraordinary light rays.
Using the same parameterization as in the previous section, one finds that
the definitions of $A$ and the polarization vector $\tilde{v}$, see~\eqref{eq:polV}, imply $F_{23}$ vanishes.
Now choose $f_{01},f_{03},f_{31}$ to be suitable multiples of a Gaussian in $\vec{k}$:
\begin{align*}
  f_{01}(\vec{k}) &= \frac{\im \tau_0^3}{\sqrt{\uppi^3} (1+\xi^2)} \exp\bigl( -(\tilde\omega \tau_0)^2 \bigr), \\
  f_{03}(\vec{k}) &= -\frac{4 \im k_1 k_3 \tau_0^5}{\sqrt{\uppi^3} (1+\xi^2)^2} \exp\bigl( -(\tilde\omega \tau_0)^2 \bigr), \\
  f_{31}(\vec{k}) &= \frac{4 \im \tilde\omega k_3 \tau_0^5}{5 \sqrt{\uppi^3} (1+\xi^2)^2} \exp\bigl( -(\tilde\omega \tau_0)^2 \bigr)
\end{align*}
and set
\begin{equation*}
  f = -f_{01} \sinh\alpha \sin\beta + f_{03} \sinh\alpha \cos\beta + f_{31} \cosh\alpha
\end{equation*}
Then we compute
\begin{equation*}
  F_{02}(0) = \im \int_{\RR^3} k_1 k_2 f(\vec{k})\, \dif\vec{k} = 0,
  \quad
  F_{12}(0) = \im \int_{\RR^3} \tilde\omega k_2 f(\vec{k})\, \dif\vec{k} = 0.
\end{equation*}
and the only non-zero components are
\begin{align*}
  F_{01}(0) &= -\im \int_{\RR^3} (\tilde\omega^2 - k_1^2) f(\vec{k})\, \dif\vec{k} = -\tau_0^{-2} \sinh\alpha \sin\beta, \\
  F_{03}(0) &= \im \int_{\RR^3} k_1 k_3 f(\vec{k})\, \dif\vec{k} = \tau_0^{-2} \sinh\alpha \cos\beta, \\
  F_{31}(0) &= -\im \int_{\RR^3} \tilde\omega k_3 f(\vec{k})\, \dif\vec{k} = \tau_0^{-2} \cosh\alpha.
\end{align*}
Moreover, note that $F_{ab}$ is smooth with Schwartz class components.

With this choice of $F_{ab}$ we find that $\mathfrak{b}_2^{ab} F_{ab} = 2\tau_0^{-2}$ is the only non-zero contraction of $F_{ab}$ with any of the basis vectors in~\eqref{eq:subspace_basis}.
Therefore
\begin{equation*}
  \rho(0) = 4 (1 - \xi^2 \sinh^2\alpha \sin^2\beta) \tau_0^{-4} < 0
\end{equation*}
in the interluminal case.
Since $\rho$ is smooth it follows that $\rho < 0$ in an open neighbourhood of the origin.

In summary, we have shown that the classical sWEC is violated for interluminal observers (in certain positive frequency complex-valued solutions) and, consequently, there exists no finite lower bound for the normal ordered quantized energy density along their worldline.
The same result evidently holds for the magnetic part of the energy density by itself.
However, the electric part is positive definite and so we cannot conclude from these arguments whether or not it is unbounded below in the QFT. Further insight might be gained by considering states that are superpositions of the vacuum with a two-particle state, as in~\cite{FordHelferRoman:2002,FewsterRoman:2003}.

%******************************************************************************%
\section{Discussion}

The main result of this article is the rigorous derivation of a state-independent Quantum Energy Inequality for certain types of observers in pre-metric linear electrodynamics, and its explicit calculation in the illustrative and physically interesting example of a uniaxial crystal.
This required a classification of possible observer trajectories (extending previous work by~\cite{Raetzel:2010je}) to account for the richer causal structure possible in the pre-metric theory, compared to the usual Lorentzian metric structure of Maxwell electrodynamics.
For reduced, bihyperbolic, energy-distinguishing and time-distinguishing Fresnel polynomials, we classified future-pointing trajectories as either sub-, inter- and superluminal, depending on the relation of their tangents to the null structure of the dual polynomial.
Such a classification is unnecessary in Lorentzian geometry, where only one class of future-pointing trajectories exists, namely timelike trajectories.

The clarification of possible observer trajectories set the language to discuss QEIs for quantized pre-metric electrodynamics, which we proved to hold on general grounds for subluminal observers in Sect.~\ref{sub:qei:proof}.
In Sect.~\ref{sec:uniaxial}, we derived an explicit QEI bound for subluminal observers moving at uniform velocity relative to the medium.
We were particularly careful about the normalization of the observer trajectories which may differ according to the interpretation of the uniaxial electrodynamics model.
While the value of the QEI bound depends on the normalization chosen, its divergence at the extraordinary lightcone does not.
To gain an insight into the quantized energy density along non-subluminal directions, we also showed in Sect.~\ref{sub:uniaxial-qei:interluminal} that there exist quantum states in which the energy density can become arbitrarily negative (independent of normalization).

The next steps in the quantization of pre-metric electrodynamics are the rigorous construction of the quantized theory for non-constant constitutive laws and its coupling to other fields.
Since general non-constant constitutive laws can lead to lightcone structures which split, combine and cross, not only is the causal behavior of such theories more complicated but also the construction of propagators faces additional difficulties.
For instance, the problem of propagation of singularities for distributional solutions to PDEs with lightcones of variable multiplicity has not been conclusively solved in the mathematical literature (see, \eg~\cite{Dencker1992,Nolan:2007}).

Regarding the coupling to other fields, a first step towards a spinor theory on a background geometry determined by Fresnel polynomial has been made in~\cite{Grosse-Holz:2017rdt}.
Next one could attempt a quantization (in the algebraic approach) of the spinor theory based on a spacetime dependent Fresnel polynomial.
It would be interesting to investigate QEIs for such a theory (see~\cite{FewVer:2002,DawFew:2006} for general QEIs on Dirac fields in the metric case).

An interesting phenomenon which appears also in quantized pre-metric electrodynamics is the Casimir effect.
In~\cite{Rivera:2011rx}, the Casimir effect was already studied in media with a certain bi-metric Fresnel polynomial.
Based on the results presented in this article, the Casimir effect could be investigated explicitly in uniaxial crystals and a priori bounds for other media could be given.

Apart from being interesting in their own right, non-linear media have a promising application as analog models for quantum gravity.
For example, as in~\cite{FordLorenci1,FordLorenci2,FordLorenci3} one can study non-linear dielectrics as an analog model for lightcone fluctuations.
The pre-metric approach might provide a clearer conceptual footing to this problem.

%------------------------------------------------------------------------------%
\begin{acknowledgments}
  The authors thank Frederic Schuller for insightful discussions, Markus Fr\"ob for discussions on Appendix~\ref{appx:int}, and Paul Busch for some useful remarks on the manu\-script.
  The work of D.S.\ was supported by a grant of the Polish National Science Center (NCN) based on the decision no.~DEC-2015/16/S/ST1/00473.
  C.P.\ gratefully thanks the Center of Applied Space Technology and Microgravity (ZARM) at the University of Bremen for their kind hospitality and acknowledges support of the European Regional Development Fund through the Center of Excellence TK133 ``The Dark Side of the Universe''.
  C.J.F.\ thanks the Department of Mathematical Methods in Physics at the University of Warsaw for its warm hospitality during a visit in the course of this work.
\end{acknowledgments}

%------------------------------------------------------------------------------%
\bigskip
\appendix

\section{Identities used to evaluate the quantized point-split energy density}\label{appx:int}

In this Appendix, we prove the identities~\eqref{eq:abk} and~\eqref{eq:abkt} used in Sect.~\ref{sec:uniaxial}.

Let $f\in C_0^\infty(\mathbb{R})$ and let $u$ and $v$ be fixed $4$-vectors with $u$ $\eta$-timelike and future-pointing.
We claim that
\begin{align}\label{eq:identity}
  \frac{1}{(2\uppi)^3}  \int_{\RR^3} \frac{(k\cdot u)( k\cdot v)}{2\omega}  \what{f}(k\cdot u) \,\dif\vec{k} &=
  -\frac{\eta(u,v)}{4\uppi^2 \eta(u,u)^2}
  \int_0^\infty \kappa^{3}\what{f}(\kappa)\,\dif\kappa.
\end{align}
To prove this, first observe that if $(k\cdot u)( k\cdot v)$ in~\eqref{eq:identity} were replaced by $k_a k_b$, the resulting integral would be constructed covariantly from $\eta_{ab}$ and $u_a u_b$ and would vanish on contraction with $\eta^{ab}$; it is therefore proportional to $\eta_{ab}-4 u_a u_b/\eta(u,u)$.
Therefore the left-hand side of~\eqref{eq:identity} equals $A \eta(u,v)$, where the constant of proportionality $A$ is fixed by the special case $v=u$.
Using Lorentz-invariance of the measure $(2\omega)^{-1}\dif\vec{k}$, this integral may be evaluated in the rest-frame of $u$, whereupon
\begin{equation*}
  A\eta(u,u)
  = -\frac{\eta(u,u)}{2(2\uppi)^3}\int_{\RR^3} \omega \what{f}(\omega\sqrt{-\eta(u,u)})\, \dif\vec{k}
  = -\frac{\eta(u,u)}{4\uppi^2}\int_0^\infty \omega^3 \what{f}(\omega\sqrt{-\eta(u,u)})\, \dif\omega.
\end{equation*}
Changing variables to $\kappa = \omega\sqrt{-\eta(u,u)}$ gives the required result~\eqref{eq:identity}.

Removing the test function, \eqref{eq:identity} implies
\begin{equation}\label{eq:abk}
  \frac{1}{(2\uppi)^3} \int_{\RR^3} \frac{(k\cdot u) (k\cdot v)}{2\omega}\e^{-\im k\cdot u (\tau-\tau')}\, \dif\vec{k} = -\frac{\eta(u,v)}{4\uppi^2\eta(u,u)^2} \int_0^\infty \kappa^{3}\e^{-\im\kappa (\tau-\tau')}\, \dif\kappa.
\end{equation}
The above derivation applies equally well if $\eta$ is replaced by $\zeta$ and $k$ by $\tilde{k}$, so we also have
\begin{equation}\label{eq:abkt}
  \frac{1}{(2\uppi)^3} \int_{\RR^3} \frac{(\tilde{k}\cdot u)(\tilde{k}\cdot v)}{2\tilde{\omega}} \e^{-\im \tilde{k}\cdot u (\tau-\tau')} \,\dif\vec{k} = -\frac{\zeta(u,v)}{4\uppi^2\zeta(u,u)^2} \int_0^\infty \kappa^{3}\e^{-\im\kappa (\tau-\tau')}\,\dif\kappa,
\end{equation}
for any $\zeta$-timelike $4$-vector $u$.
(Alternatively, one can make a change of variables to reduce the left-hand side to another instance of \eqref{eq:abk}.)

%------------------------------------------------------------------------------%
\section{The normal ordered energy density acting on the vacuum}
\label{appx:ReehSchlieder}

Let $f$ be a smooth compactly supported real-valued function. We aim to exclude
the possibility that $\norder{\what\rho(f)} \Omega=0$ unless $f$ vanishes identically -- a result analogous to an instance of the Reeh--Schlieder theorem, needed for our argument in Sect.~\ref{sub:uniaxial-qei:neg} on the existence of states in which the expectation value of the quantized energy density is negative.
Writing $\norder{\what\rho(f)} \Omega$ out, we find
\begin{align*}
  \norder{\what\rho(f)} \Omega
  &= -\frac{\varepsilon(e)^{-1}}{2(2\uppi)^6} \bigl(\chi^{abcd} - 2\dot\gamma^a n_e \chi^{ebcd} - 2\dot\gamma^c n_e \chi^{abed} \bigr) \\&\quad\times \iint_{\RR^3 \times \RR^3} \frac{k_{[a} v_{b]} k'_{[c} v'_{d]}}{\sqrt{4\omega\omega'}} \what{f}(k+k') a^*(\vec{k}) a^*(\vec{k}')\, \dif\vec{k}\, \dif\vec{k}'\, \Omega + \text{other terms},
\end{align*}
where the other terms lie in orthogonal subspaces of the $2$-particle subspace, generated by creation operators such as $\tilde{a}^*(\vec{k}) a^*(\vec{k}')$ or $\tilde{a}^*(\vec{k}) \tilde{a}^*(\vec{k}')$.

One calculates
\begin{equation*}
  \norder{\what\rho(f)} \Omega = -\frac{\varepsilon(e)^{-1}}{(2\uppi)^6} \iint_{\RR^3 \times \RR^3} r(\vec{k},\vec{k}') \what{f}(k+k')  a^*(\vec{k}) a^*(\vec{k}')\, \dif\vec{k}\, \dif\vec{k}'\, \Omega + \text{other terms},
\end{equation*}
where $r(\vec{k},\vec{k}')$ is given by
\begin{align*}
  4\sqrt{\omega\omega'} r(\vec{k}, \vec{k}') &=
  \eta^{-1}(v',k) \bigl( \eta^{-1}(v,n) (k' \cdot \dot\gamma) + \eta(k',n) (v \cdot \dot\gamma) \bigr) \\&\quad
  + \eta^{-1}(v,k') \bigl( \eta^{-1}(v',n) (k \cdot \dot\gamma) + \eta(k,n) (v' \cdot \dot\gamma) \bigr) \\&\quad
  - \eta^{-1}(v,v') \bigl( \eta^{-1}(k',n) (k \cdot \dot\gamma) + \eta(k,n) (k' \cdot \dot\gamma) \bigr) \\&\quad
  - \eta^{-1}(k,k') \bigl( \eta^{-1}(v',n) (v \cdot \dot\gamma) + \eta^{-1}(v,n) (v' \cdot \dot\gamma) \bigr) \\&\quad
  - \eta^{-1}(v,k') \eta^{-1}(v',k)
  + \eta^{-1}(k,k') \eta^{-1}(v,v')
\end{align*}
and the `other terms' have a similar form.
Let $k = k'$.
Then, using $\eta^{-1}(k,k)=0$, $\eta^{-1}(v,k)=0$, $\eta^{-1}(v,v)=1$,
\begin{equation*}
  r(\vec{k}, \vec{k}) = -\frac{\eta^{-1}(k,n) (k \cdot \dot\gamma)}{2\omega}.
\end{equation*}
Noting that $k \cdot\dot\gamma > 0$ and $\eta^{-1}(k,n) < 0$, we find $r(\vec{k}, \vec{k}) > 0$.
Therefore there is a nonempty open set $N\subset\mathbb{R}^3\times\mathbb{R}^3$ on which $\inf_N r> 0$.

It is clear that
\begin{equation*}
  \norm{\norder{\what\rho(f)} \Omega}^2 \ge \frac{2}{(2\uppi)^6}\iint_{N} |r(\vec{k},\vec{k'})|^2 |\what{f}(k+k')|^2 \,\dif\vec{k}\, \dif\vec{k}',
\end{equation*}
and it follows that $\norder{\what\rho(f)} \Omega =0$ only if $\what{f}$ vanishes almost everywhere on $\{k+k'\mid (\vec{k},\vec{k'})\in N\}$, which has nonempty interior in $\mathbb{R}^4$. As $f$ is compactly supported, $\what{f}$ is analytic, and we may conclude that it (and hence $f$) vanishes identically.

%------------------------------------------------------------------------------%
\section{The dual Lagragian for massive momenta}
\label{appx:MassiveDual}

To derive the intrinsic normalization factor $\N_{\mathrm{UC}}$ for the curves employed in Sect.~\ref{sub:uniaxial-qei:energy}, we follow \cite{Raetzel:2010je}.
We derive the dual Lagrangian determining the trajectories of particles and observers with massive momenta via the Legendre map from the dispersion relation.
In our case the dispersion relation is defined by the Fresnel polynomial of the crystal
\begin{equation*}
  P(k) = |\eta|^{-1/2}\mathcal{G}(k) = \eta^{-1}(k,k)\zeta^{-1}(k,k) = m^4,
\end{equation*}
see \ref{eq:Fresnel-uni}.
Since this calculation is rather involved, we perform the derivation only up to third order in the crystal parameter $\xi$.

The starting point is the Helmholtz action for free particles satisfying the massive dispersion relation
\begin{align*}
	S[x,k,\lambda] = \int \Bigl(k \cdot \dot{x} - \lambda \ln\bigl(P(\tfrac{k}{m})\bigr) \Bigr)\, \dif\tau,
\end{align*}
where variation with respect to $\lambda$ enforces the dispersion relation $\mathcal{P}(k) = m^4$.
By successive variation with respect to $k$ and $\lambda$, it is possible to remove the dependence of the action on $k$ and $\lambda$ to obtain an action which determines the motion of massive point particles and the proper time of observer clocks
\begin{equation*}
	S[x] = m \int P^*(\dot x)\, \dif\tau,
\end{equation*}
where $P^*$ is a one-homogeneous function with respect to $\dot x$.

In \cite{Raetzel:2010je} it was shown that
\begin{equation*}
	P^*(\dot x) = P( k(\dot x))^{-\frac{1}{4}}.
\end{equation*}
The term $k(\dot x)$ is the inverse of the Legendre map
\begin{align*}
	\dot x^a (k)
	&= \frac{1}{4}\frac{\partial_{k_a}\mathcal{G}(k)}{\mathcal{G}(k)} = \frac{1}{2}\frac{k^a}{\eta^{-1}(k,k)} + \frac{1}{2} \frac{(\zeta^{-1})^{ab}k_b}{\zeta^{-1}(k,k)} \\
	&= \frac{k^a}{\eta^{-1}(k,k)} + \frac{\bigl( \xi^2 (k \cdot U)^2 - (k \cdot X)^2 \bigr)\, k^a}{2 \eta^{-1}(k,k)^2} \\&\quad + \frac{\bigl( (k \cdot X) X^a - \xi^2 (k \cdot U) U^a \bigr)}{2 \eta^{-1}(k,k)} + \mathcal{O}(\xi^4)
\end{align*}
which we expanded up to $\xi^2$ (the term of order $\xi^3$ vanishes).
Up to this order, the inverse of the Legendre map is given by
\begin{align*}
	k_a(\dot x) &= \frac{\dot x_a}{\eta(\dot x, \dot x)} + \frac{\bigl(\eta(\dot x, X)^2 - \xi^2 \eta(\dot x, U)^2 \bigr)\, \dot x_a}{2 \eta(\dot x, \dot x)^2} \\&\quad + \frac{\bigl( \xi^2 \eta(\dot x, U)U_a - \eta(\dot x, X) X^a \bigr)}{2 \eta(\dot x,\dot x)} + \mathcal{O}(\xi^4).
\end{align*}

Using this expression, we find
\begin{equation*}
	P^*(\dot x) = \sqrt{\eta(\dot x, \dot x)} + \frac{\xi^2 \eta(\dot x,U)^2 - \eta(\dot x, X)^2}{4 \sqrt{\eta(\dot x,\dot x)}} + \mathcal{O}(\xi^4).
\end{equation*}
Thus an observer curve $\gamma$ is uniaxial crystal electrodynamics proper time parametrized if and only if $P^*(\dot\gamma) = 1$.

For the worldline \eqref{eq:worldline} along which we study the point-split energy density in Sect.~\ref{sub:uniaxial-qei:energy} we find
\begin{equation*}
	P^*(\dot\gamma) = \N_{\mathrm{UC}} \biggl( 1 + \frac{\xi^2}{4} (1 + \sinh^2\alpha \sin^2\beta) \biggr) + \mathcal{O}(\xi^4)
\end{equation*}
and thus the normalization factor must be
\begin{equation}\label{eq:uniaxnorm}
	\N_{\mathrm{UC}} = 1 - \frac{\xi^2}{4}(1 + \sinh^2\alpha \sin^2\beta) + \mathcal{O}(\xi^4).
\end{equation}

%------------------------------------------------------------------------------%

\small
\bibliographystyle{cmphref}
\bibliography{QEIinGLED.bib}

\end{document}